\documentclass[11pt,a4paper]{article}
\usepackage{graphicx}
\usepackage{amssymb}
\usepackage{amsmath}
\usepackage{braket}
\usepackage{listings}
\usepackage{subfig}
\usepackage{cases}
\usepackage{comment}
\usepackage{ltgtsim}
\usepackage{jcappub}
\newcommand{\lambdabar}{\lambda \kern -0.5em\raise 0.5ex \hbox{--}}
\newcommand{\dif}[2]{\frac{d #1}{d #2}}

\def\slashchar#1{\setbox0=\hbox{$#1$} 
\dimen0=\wd0 
\setbox1=\hbox{/} \dimen1=\wd1 
\ifdim\dimen0>\dimen1 
\rlap{\hbox to \dimen0{\hfil/\hfil}} 
#1 
\else 
\rlap{\hbox to \dimen1{\hfil$#1$\hfil}} 
/ 
\fi}

\newtheorem{definition}{Definition}[section]

\title{Non-perturbative approach for curvature perturbations in stochastic-\boldmath$\delta N$ formalism}

\author[a,b]{Tomohiro Fujita,}
\author[a,c]{Masahiro Kawasaki}
\author[a,b,d]{and Yuichiro Tada}
\affiliation[a]{Kavli Institute for the Physics and Mathematics of the
Universe (WPI), TODIAS,  the University of Tokyo, 5-1-5
Kashiwanoha, Kashiwa, 277-8583, Japan}
\affiliation[b]{Department of Physics, the University of Tokyo, Bunkyo-ku
113-0033, Japan}
\affiliation[c]{Institute for Cosmic Ray Research, the University of Tokyo,
5-1-5 Kashiwa-no-Ha, Kashiwa, Chiba, 277-8582, Japan}
\affiliation[d]{Advanced Leading Graduate Course for Photon Science (ALPS), the University of Tokyo, Bunkyo-ku
113-0033, Japan}

\emailAdd{tomohiro.fujita@ipmu.jp}
\emailAdd{kawasaki@icrr.u-tokyo.ac.jp}
\emailAdd{yuichiro.tada@ipmu.jp}

\abstract{In our previous paper~\cite{Fujita:2013cna}, we have proposed a new algorithm to calculate the power spectrum of the curvature perturbations
generated in inflationary universe
with use of the stochastic approach. Since this algorithm does not need the perturbative expansion with respect to the inflaton 
fields on super-horizon scale, 
it works even in highly stochastic cases.  
For example, when the curvature perturbations are very large
or the non-Gaussianities of the curvature perturbations are sizable,
the perturbative expansion may break down but 
our algorithm enables to calculate the curvature perturbations.
We apply it to two well-known inflation models, chaotic and hybrid inflation, 
in this paper. 
Especially for hybrid inflation, 
while the potential is very flat around the critical point
and the standard perturbative computation is problematic,
we successfully calculate the curvature perturbations.}


\arxivnumber{1405.2187}

\begin{document}

\begin{flushright}
ICRR-Report-680-2014-6
\\
IPMU 14-0113
\end{flushright}

\maketitle
\flushbottom

\section{Introduction}
Recently, the BICEP2 collaboration has discovered the primordial B-mode polarization of the cosmic microwave background 
(CMB)~\cite{Ade:2014xna}.
This strongly supports the inflationary paradigm, the accelerating expansion in the early universe.
In the standard inflationary paradigm, all of the 
observed fluctuations including the temperature anisotropy of CMB and the seeds of the large scale structures are assumed to originate in 
the quantum fluctuations of the inflaton, the scalar field which drives inflation. 

Though the primordial curvature perturbations generated during 
inflation are quite small $\sim10^{-5}$ on the CMB scale~\cite{Ade:2013zuv}, 
they are not necessarily to be so on smaller scales and there may exist large curvature perturbations
which lead to formation of curious astronomical objects like primordial black holes 
(PBHs)~\cite{Carr:1975qj,Carr:1976zz,Carr:2003bj} and ultracompact minihalos (UCMHs)~\cite{Josan:2010vn,Bringmann:2011ut,Li:2012qha}.
Moreover as we will mention later, hybrid inflation can generate large curvature perturbations with
a peak profile around the scale corresponding to the critical point
because the inflaton potential is very flat around that point. Therefore it is quite 
interesting to consider large curvature perturbations.

The generated curvature perturbations are often calculated by a perturbative approach with respect to the inflaton field $\phi$.
However, this approximation will not be good to analyze large curvature perturbations. That is because the effects of the higher-order
perturbations may not be negligible in such a case.
In fact, in multi-field inflation,
a field fluctuation can become so large 
compared to its homogeneous value
that the perturbative expansion breaks down.
In such a case, we need the non-perturbative approach.

In our previous paper~\cite{Fujita:2013cna}, we proposed a non-perturbative method which combined the stochastic 
approach~\cite{Starobinsky:1986fx,Starobinsky:1994bd,Sasaki:1987gy,Nakao:1988yi,Nambu:1988je,Mollerach:1990zf,Finelli:2008zg,Finelli:2010sh,Sanchez:2012tk,Sanchez:2013zst,Enqvist:2011pt,Kawasaki} and the $\delta N$ formalism~\cite{Starobinsky:1986fxa,Salopek:1990jq,Sasaki:1995aw,Sasaki:1998ug,Lyth:2004gb}.
The inflaton field which is coarse-grained over a super-horizon scale
is considered in the stochastic approach.
Since fluctuation modes of the inflaton which cross the horizon and are classicalized
give contributions to the coarse-grained field continuously,
their effects are taken into the equation of motion (e.o.m.) as statistical random noise.
The duration of inflation, namely e-folds $N$, fluctuates because of this noise, and our method connects it to the gauge-invariant
curvature perturbations $\zeta$ with use of the $\delta N$ formalism 
in a non-perturbative manner.\footnote{To see other methods which calculate the curvature perturbations with use of the stochastic approach, 
see~\cite{Kunze:2006tu,Ivanov:1997ia,Yokoyama:1998pt,Saito:2008em,Levasseur:2013ffa,Levasseur:2013tja,Enqvist:2008kt}}

In this paper, in order to show validity of our method (we call ``the stochastic-$\delta N$ formalism"),
we apply it to two inflation models, chaotic inflation~\cite{Linde:1983gd} and 
hybrid inflation~\cite{Linde:1993cn,Copeland:1994vg}. In the latter case, 
multiple fields are involved and a high stochasticity is realized.
In particular, we successfully calculate the 
curvature perturbations generated around the critical point and during the waterfall phase in hybrid infaltion
for the parameters where the waterfall phase continues more than 10 e-folds.

It should be noted that the original type of hybrid inflation is rejected by the observation of CMB by the Planck collaboration~\cite{Ade:2013zuv}
because this model predicts a blue-tilted spectrum. Moreover, the recent report of B-mode detection by the BICEP2 collaboration suggests the 
large-field (or super-Planckian-field) inflation models, though hybrid inflation is generally a small-field model.\footnote{It should 
be noted that the extended hybrid models which predict red-tilted spectra consistent with CMB observations 
are studied well~\cite{Clesse:2008pf,Clesse:2010iz}. 
Furthermore, there is a tension between Planck and BICEP2 results.}
However we don't describe these topics in detail in this paper. Instead, we consider hybrid inflation just as a toy model of multi-field inflation to show how to use the stochastic-$\delta N$ method.

The rest of the paper is organized as follows. In section~\ref{lin pert th}, we quickly review the standard linear perturbation theory.
In section~\ref{stochastic delta N}, we explain the
stochastic-$\delta N$ formalism briefly. In section~\ref{chaotic}, we demonstrate
the stochastic-$\delta N$ in chaotic inflation, and then, in section~\ref{hybrid},
we calculate the power spectrum of the curvature perturbations in hybrid inflation. 
Finally section~\ref{conclusion} is devoted to conclusion.

\section{Linear perturbation theory}\label{lin pert th}
First let us review the standard linear perturbation theory for the comparison of the stochatic-$\delta N$.

According to the Einstein equation, an accelerating expansion of a space-time can be brought about by 
the potential energy 
of a homogenous scalar field (called ``inflaton field"). 
If the inflaton field slowly rolls down on its potential, inflation can continue for a long time. 
In the isotropic and homogenous FLRW space-time,
\begin{eqnarray}
        ds^2=-dt^2+a^2(t)d\mathbf{x}^2,
\end{eqnarray}
the e.o.m. of the scalar field $\phi$ is given by
\begin{eqnarray}\label{scalar eom}
        \ddot{\phi}+3H\dot{\phi}-a^{-2}\nabla^2\phi+V_\phi=0.
\end{eqnarray}
Here $H=\dot{a}/a$ is the Hubble parameter, a dot represents a time derivative and $V_\phi$ denotes a partial derivative $\partial V/\partial\phi$. 
For a slow-rolling, $\ddot{\phi}\ll V_\phi$, and homogenous scalar field, one can obtain 
\begin{eqnarray}\label{slow-roll eq}
        3H\dot{\phi}\simeq-V_\phi.
\end{eqnarray}
If the inflaton rolls down so slowly that the kinetic energy $\dot{\phi}^2/2$ can be neglected compared to the potential energy $V$, the Friedmann 
equation leads a nearly constant Hubble parameter $H\simeq\sqrt{V/3M_p^2}$ and then the scale factor $a(t)$ grows exponentially 
$a(t)\propto e^{\int Hdt}$. This exponent part $N=\int Hdt$ is called e-folds and often used as a dimensionless time variable.

To make the slow-roll condition clear, the following slow-roll parameters are often used:
\begin{eqnarray}\label{slow-roll parameter}
        \epsilon_\phi=\frac{M_p^2}{2}\left(\frac{V_\phi}{V}\right)^2, \quad \eta_{\phi\phi}=M_p^2\frac{V_{\phi\phi}}{V},
\end{eqnarray}
where $M_p$ denotes the reduced Planck mass $\sqrt{\frac{1}{8\pi G}}\simeq2.4\times 10^{18}\,\mathrm{GeV}$.
Then the slow-roll condition is given by $\epsilon_\phi\ll1,|\eta_\phi|\ll1$.

The inflaton field is decomposed into the homogenous part and the perturbation part:
\begin{eqnarray}
        \phi(t,\mathbf{x})=\phi_0(t)+\delta\phi(t,\mathbf{x}).
\end{eqnarray}
Assuming the perturbation $\delta\phi$ is much smaller than the zero mode $\phi_0$, 
the linearized e.o.m. for the Fourier mode $\phi_\mathbf{k}$ is obtained from eq.~(\ref{scalar eom}) as
\begin{eqnarray}\label{lin eom}
        \ddot{\phi}_\mathbf{k}+3H\dot{\phi}_\mathbf{k}+\left(\frac{k^2}{a^2}+V_{\phi\phi}(\phi_0)\right)\phi_\mathbf{k}=0.
\end{eqnarray}
By approximating $V_{\phi\phi}$ by a constant mass $m^2$ and adopting the Bunch-Davies vacuum 
as the initial condition of inflation, one finds the solution of this equation 
as
\begin{eqnarray}
        \phi_\mathbf{k}=\frac{\sqrt{\pi}}{2}H\left(\frac{1}{aH}\right)^{3/2}H_\nu^{(1)}\left(\frac{k}{aH}\right),
\end{eqnarray}
where $H_\nu^{(1)}$ is the Hankel function of the first kind and $\nu$ is defined as
\begin{eqnarray}
        \nu=\sqrt{\frac{9}{4}-\frac{m^2}{H^2}}\simeq\frac{3}{2}-\frac{m^2}{3H^2}.
\end{eqnarray}
Here the inflaton mass $m$ should be negligible compared to the Hubble parameter for slow-roll inflation.
One can obtain the power spectrum 
which is the two-point correlator of the inflaton field
in Fourier space 
as
\begin{eqnarray}\label{const mass power}
        \mathcal{P}_\phi(k)&=&\frac{k^3}{2\pi^2}\int d^3x\braket{\phi(\mathbf{x}=0)\phi(\mathbf{x})}e^{-i\mathbf{k}\cdot\mathbf{x}} \nonumber \\
        &=&\frac{k^3}{2\pi^2}|\phi_\mathbf{k}|^2=\frac{H^2}{8\pi}\left(\frac{k}{aH}\right)^3\left|H_\nu^{(1)}\left(\frac{k}{aH}\right)\right|^2.
\end{eqnarray}  
With use of the asymptotic form of the Hankel function,
\begin{eqnarray}\label{asymptote of Hankel}
        H_\nu^{(1)}(x)\to-i\frac{\Gamma(\nu)}{\pi}\left(\frac{2}{x}\right)^\nu, \quad \text{$\mathrm{Re}\nu>0$ and $x\to+0$,} 
\end{eqnarray}
it is shown that the power spectrum gets \emph{frozen} to a constant on the super-horizon scale,
\begin{eqnarray}
        \mathcal{P}_\phi(k)\to\left(\frac{H}{2\pi}\right)^2, \quad \frac{k}{aH}\to0,
\end{eqnarray}

The perturbations of the duration of inflation due to this frozen quantum fluctuations cause the metric curvature perturbations. 
In fact the scale factor, which is the spatial part of the metric, is proportional to $e^N$, and therefore
the fluctuation of e-folds $\delta N$ is nothing but the metric perturbation. 
According to the $\delta N$ formalism~\cite{Starobinsky:1986fxa,Salopek:1990jq,Sasaki:1995aw,Sasaki:1998ug,Lyth:2004gb}, the gauge-invariant
curvature perturbation $\zeta$ can be calculated up to the first order perturbation of $\phi$ as
\begin{eqnarray}\label{linear delta N}
        \zeta=\dif{N}{\phi}(\phi)\delta\phi,
\end{eqnarray}
where $N(\phi)$ denotes the e-folds taken from $\phi$ to the inflation end value $\phi_f$ and can be obtained from the 
slow-roll eq.~(\ref{slow-roll eq}) as
\begin{eqnarray}
        N(\phi)=-\int^{\phi_f}_\phi\frac{V}{V_\phi M_p^2}d\phi.
\end{eqnarray}
Thus we obtain the standard result on the power spectrum of the curvature perturbations as
\begin{eqnarray}\label{linear result}
        \mathcal{P}_\zeta(k)=\left.\left(\frac{V}{V_\phi M_p^2}\right)^2\mathcal{P}_\phi\right|_{k=aH}=\left.\frac{1}{24\pi^2M_p^4}\frac{V}{\epsilon_\phi}
        \right|_{k=aH}.
\end{eqnarray}
In this paper, we demonstrate the numerical calculations of the stochastic-$\delta N$ approach, which is more general and efficient algorithm
especially when the perturbative expansion~(\ref{linear delta N}) is broken down.

\section{\texorpdfstring{Stochastic-$\delta N$ formalism}{Stochastic-delta N formalism}}\label{stochastic delta N}
We briefly describe the stochastic formalism~\cite{Starobinsky:1986fx,Starobinsky:1994bd,Sasaki:1987gy,Nakao:1988yi,Nambu:1988je,Mollerach:1990zf,Finelli:2008zg,Finelli:2010sh,Sanchez:2012tk,Sanchez:2013zst,Enqvist:2011pt,Kawasaki} and 
our algorithm~\cite{Fujita:2013cna} in this section.
In the stochastic formalism, not the homogenous field but the super-horizon scale coarse-grained field is treated as the background field.
In this paper, we call this coarse-grained field the IR part which can be defined as
\begin{eqnarray}\label{def of IR}
        \phi_\mathrm{IR}(\mathbf{x},t) = \int\frac{d^3k}{(2\pi)^3}\theta(\epsilon a(t)H(t)-k)\phi_\mathbf{k}(t)e^{-i\mathbf{k}\cdot\mathbf{x}}.
\end{eqnarray}
Here $\theta$ denotes the step function and
$\epsilon$ is a positive constant parameter. Due to the step window function in the eq.~(\ref{def of IR}), 
the IR part contains only $k<\epsilon aH$ modes. 
With tiny $\epsilon$, 
wavelengths in the IR part are much longer than the horizon scale $(aH)^{-1}$.
In this paper, we set this $\epsilon$ parameter to $0.01$.

The IR part is assumed to be a classical field, and since the horizon scale $(aH)^{-1}$ becomes shorter and shorter,
the sub-horizon modes come into the IR part and get \emph{classicalized} successively. 
At this time, the field value of that classicalized mode follows
the Gaussian distribution whose variance is equal to the power spectrum.
Because of this effect, the IR part follows the Langevin equation, which is the equation of motion with white noise.
Taking account of only the mass term in the potential for sub-horizon modes, the e.o.m. of the IR part is written as~\cite{Morikawa:1989xz},
\begin{eqnarray}\label{Langevin eom}
        \begin{cases}
                \displaystyle
                \dot{\phi} = \pi + \mathcal{P}_\phi^{1/2}H^{1/2}\xi_R, \\
                \displaystyle
                \dot{\pi} = -3H\pi +a^{-2}\nabla^{2}\phi -V_\phi +q_R\mathcal{P}_\phi^{1/2}H^{1/2}\xi_R+q_I\mathcal{P}_\phi^{1/2}H^{1/2}\xi_I,
        \end{cases}
\end{eqnarray}
where $\xi_R$ and $\xi_I$ represent the white noise and $q_R$ and $q_I$ are the real and imaginary part of the following function 
$q_\nu(\epsilon)$,
\begin{eqnarray}\label{q_nu}
        q_\nu(\epsilon) = -H\left(\frac{3}{2}-\nu+\epsilon\frac{H_{\nu-1}^{(1)}(\epsilon)}{H_\nu^{(1)}(\epsilon)}\right).
\end{eqnarray}
We will describe these terms in detail below.
Note that we omit the subscript IR for simplicity.

The terms of $\xi_R$ and $\xi_I$ denote the effect that the mode crossing the horizon joins in the IR part, and
without these terms, the eqs.~(\ref{Langevin eom}) coincides with eq.~(\ref{scalar eom}).
$\xi_R$ and $\xi_I$ correspond to the classicalizations of the $\phi$ and its momentum conjugate.
However since the true conjugate is not $\dot{\phi}$ but the conformal time derivative of $a\phi$, both of $\xi_R$ and $\xi_I$ 
contribute to the dynamics of $\pi$.
$\xi_R$ and $\xi_I$ are independent zero-mean Gaussian random variables and their amplitudes are renormalized as follows.
\begin{eqnarray}
        \begin{cases}
                \displaystyle
                \braket{\xi_R(\mathbf{x},t)\xi_R(\mathbf{x}^\prime,t^\prime)} = \braket{\xi_I(\mathbf{x},t)\xi_I(\mathbf{x}^\prime,t^\prime)}
                = \frac{\sin(\epsilon aHr)}{\epsilon aHr}\delta(t-t^\prime), & r = |\mathbf{x}-\mathbf{x}^\prime|,\\
                \displaystyle
                \braket{\xi_R(\mathbf{x},t)\xi_I(\mathbf{x}^\prime,t^\prime)} = 0.
        \end{cases}
\end{eqnarray}
The reason why there is no correlation over different time is as follows. Since we choose the step function as the window function,  
only the mode $k=\epsilon aH$ joins to the IR part at each time. Therefore, for example, $\xi_R$ can formally be
written as
\begin{eqnarray}
        \xi_R\propto\int\frac{d^3k}{(2\pi)^3}\delta(k-\epsilon aH)\phi_\mathbf{k}e^{i\mathbf{k}\cdot\mathbf{x}}.
\end{eqnarray}
The correlator of $\xi_R$ is proportional to $\braket{\phi_\mathbf{k}\phi_{\mathbf{k}^\prime}}\propto\delta(\mathbf{k}-\mathbf{k}^\prime)$,
but due to the delta function $\delta(k-\epsilon aH)$, it is proportional to $\delta(\epsilon a(t)H(t)-\epsilon a(t^\prime)H(t^\prime))\propto
\delta(t-t^\prime)$. Similarly, $\xi_I$ also has no correlation over different time.
The spatial correlation decreases by the factor $\sin(\epsilon aHr)/\epsilon aHr$.
Since this factor is oscillating and  we are interested only in the coarse-grained field, 
it can be approximated by the step function $\theta(1-\epsilon aHr)$. In other words, the noise approximately has no correlation
over the horizon scale.

$\mathcal{P}_\phi$ is evaluated at the horizon exit $k=\epsilon aH$ in eq.~(\ref{Langevin eom}),
and the value $\left(\frac{H}{2\pi}\right)^2$ is often used, though one should be careful in the massive scalar case as we will
mention in section~\ref{hybrid}.
$q_R$ and $q_I$ are the real and imaginary part of the function $q_\nu(\epsilon)$~(\ref{q_nu}) as mentioned above.
They represent the time variation of $\mathcal{P}_\phi$. Indeed, with the slow-roll approximation $\nu\simeq3/2+\eta_{\phi\phi}$, 
it is shown that $q_\nu(\epsilon)$ is the first order of $\eta_{\phi\phi}$ 
and the second order of $\epsilon$ from eq.~(\ref{asymptote of Hankel}), and hence $q_\nu$ is negligible.   
Moreover, in a highly massive case, namely $\nu<3/2$, the power spectrum will be suppressed to $\mathcal{O}(\epsilon^{3-2\nu})$ by a steep potential,
so the $q_\nu$ terms are small in either case and we omit these terms in the numerical calculation.

\begin{figure}
        \begin{center}
                \includegraphics[width=10cm]{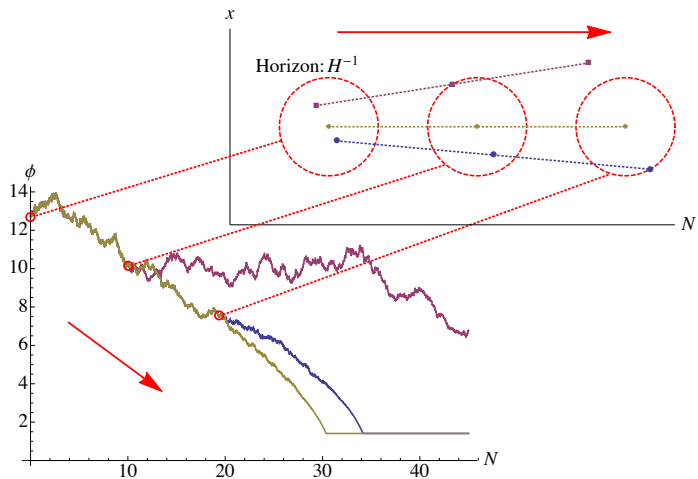}
                \caption{ 
Schematic illustration of the branching of the coarse-grained field value.
The three solid lines in the bottom graph represent $\phi_\mathrm{IR}(t)$
at three different spatial points. 
The time evolutions of these spatial points are shown as the dotted lines with corresponding colors in the above diagram. 
The red dashed circles denotes the horizon patch of the yellow point for each instant of time.
At first, $\phi_\mathrm{IR}$ at the three points develop in the same way because they are in the same Hubble patch and they receive the same white noise. However, $\phi_\mathrm{IR}$ at the magenta point first, and then 
$\phi_\mathrm{IR}$ at the blue point deviate from $\phi_\mathrm{IR}$ at the yellow point. This is because when these spatial points exit from the Hubble patch of the yellow point, their white noises and hence their time evolutions
of $\phi_\mathrm{IR}$ become independent of those at the yellow point.}
                \label{correlate}
        \end{center}
\end{figure}

In summary, the super-horizon coarse-grained field is treated as the background field in the stochastic formalism, and it follows the Langevin
eqs.~(\ref{Langevin eom}) including the white noise, $\xi_R$ and $\xi_I$. They have white spectra for time and no correlation over the horizons.
The comoving horizon scale $(aH)^{-1}$ decreases as time goes on (in other words, the physical distance $a/k$ increases), and therefore
background fields at two spatial points evolve together until their horizon crossing $k=\epsilon aH$ and then they develop independently (see figure~\ref{correlate}).
In principle, solving these Langevin eqs. over all spatial points, we can obtain the coarse-grained curvature perturbation $\zeta$.
However, it is hard to solve the Langevin eqs. for all space points simultaneously considering the branching off of each point mentioned above,
both analytically and numerically. Therefore, we use the stochastic-$\delta N$ algorithm proposed in ref.~\cite{Fujita:2013cna}. \\

The stochastic-$\delta N$ is the algorithm to calculate the curvature perturbations without a perturbative expansion with respect to the inflaton field, taking
advantage of the $\delta N$ formalism~\cite{Starobinsky:1986fxa,Salopek:1990jq,Sasaki:1995aw,Sasaki:1998ug,Lyth:2004gb}.
In the stochastic formalism, the background field evolves, receiving the horizon scale noise. Therefore, the duration of inflation for each 
Hubble patch, namely e-folds $N$, is automatically fluctuated. According to the $\delta N$ formalism, these fluctuations of e-folds
$\delta N$ are nothing but the gauge invariant curvature perturbations $\zeta$. The power spectrum of the curvature perturbations is just the correlator of 
$\delta N$.

The key obstacle to calculate the perturbations is, 
as mentioned above, the difficulties of solving the Langevin eqs. over the all spatial points. On the contrary,
the evolution at one space point can be calculated easily by a numerical simulation. The stochastic-$\delta N$ formalism extracts the information
of correlations from the one-point evolutions cleverly.
Let us describe our algorithm below.
\begin{itemize}
\item[1.]Choose ``initial" value $\phi_i$ for the inflaton field, from which the calculation is started.\footnote{
Note that this ``initial" value is set artificially to obtain the correspondence between $\braket{N}$ and $\braket{\delta N^2}$
and does not represent the physical initial condition of inflation.
} 

\item[2.]Integrating the Langevin equations from that ``initial" value numerically, 
obtain the e-folds $N$ which the inflaton field takes until it rolls reaches $\phi_f$ where the inflation ends.\footnote{Naively, 
the point where the slow-roll condition gets violated can be chosen as $\phi_f$. 
This choice is valid if inflation is driven by a single-inflaton field. However, in multi-field inflation cases, 
a point in the field space where the slow-roll condition is violated 
is not unique and the potential energies are not necessarily same
at those points.
Since the end time slice of the $\delta N$ formalism should be an uniform density slice, the uniform Hubble slice
should be chosen instead.} Since the Langevin equations include random noise, the e-fold varies in each calculation.
Each e-fold corresponds to the duration of inflation in some Hubble patch and their fluctuations represent the super-horizon coarse-grained 
curvature perturbations. 
Therefore, reiterating the calculations, we can get the spatial mean and variance of e-folds, namely $\braket{N}$
and $\braket{\delta N^2}$.

\item[3.]Next, reiterate the above calculations changing $\phi_i$ and obtain other sets of $\braket{N}$ and $\braket{\delta N^2}$. Thus, we obtain
$\braket{\delta N^2}$ as a function of $\braket{N}$ finally.

\item[4.]Here, recall that the power spectrum of the curvature perturbations is defined as the Fourier mode of the correlator of $\delta N$
as follows.
\begin{eqnarray}
        \mathcal{P}_\zeta = \mathcal{P}_{\delta N} = \frac{k^3}{2\pi^2}\int d^3x\braket{\delta N(\mathbf{x}=0)\delta N(\mathbf{x})}e^{-i\mathbf{k}\cdot
        \mathbf{x}}.
\end{eqnarray}
Inversely, the variance of e-folds can be described as the inverse Fourier mode of the power spectrum in the limit of $\mathbf{x}\to0$.
\begin{eqnarray}\label{variance and power}
        \braket{\delta N^2} = \int^{k_f}_{k_i}\frac{dk}{k}\mathcal{P}_{\delta N}(k) \simeq \int^{\ln k_f}_{\ln k_f-\braket{N}}\mathcal{P}_{\delta N}(N)dN,
\end{eqnarray} 
with the integration between the Hubble scale at the beginning of inflation, $k_i=\epsilon aH|_i$, and that at the end of inflation, $k_f=\epsilon aH|_f$,
under the assumption that every fluctuation is made during inflation. Here we also used the approximation 
that $k_i\simeq k_fe^{-\braket{N}}$. This approximation is good
if the curvature perturbation does not exceed unity and the Hubble scale $\epsilon aH$ does not spatially fluctuate much.
Since the left-hand side of eq.~(\ref{variance and power}) is already obtained as a function of $\braket{N}$ in step 3, 
we can get the power spectrum by differentiating both sides
with respect to $\braket{N}$:
\begin{eqnarray}
        \mathcal{P}_\zeta(k) = \mathcal{P}_{\delta N}(k) = \left.\dif{}{\braket{N}}\braket{\delta N^2}\right|_{\braket{N}=\ln(k_f/k)}.
\label{Pzeta calculation}
\end{eqnarray}

\end{itemize}

In the single-field case, this procedure is enough to obtain the power spectrum and we showed analytically
that the result is consistent with that of the standard linear perturbation
theory in the slow-roll limit in the previous paper~\cite{Fujita:2013cna}. 
However, we should be careful to extend it to the multi-field case, like hybrid inflation.
If there is only one inflaton, the ``initial" value $\phi_i$ and $\braket{N}$ have one-to-one correspondence, and $\braket{\delta N^2}$ is determined
once $\phi_i$ is given.
Thus $\braket{\delta N^2}$ is uniquely given as a function of $\braket{N}$. 
However, when the inflaton field space becomes multi-dimensional, 
the one-to-one correspondence between a set of ``initial" field values 
$\{\phi_i, \psi_i, \cdots \}$ and $\braket{N}$ no longer exists because 
different sets of ``initial" values
can lead to the same value of $\braket{N}$.
Then, although $\braket{\delta N^2}$ can be still calculated for each ``initial" value, the functional form of $\braket{\delta N^2} (\braket{N})$
is not unique but depends on a trajectory in the inflaton field space where inflatons go through. 
We can also rephrase it as follows. 
Both $\braket{N}$ and $\braket{\delta N^2}$ can be computed
if an arbitrary set of ``initial" values of inflatons is given. 
Therefore one can consider that a pair of 
$\braket{N}$ and $\braket{\delta N^2}$ values is assigned to every point
in the inflaton field space like potentials.
In single field cases, the field space is one-dimensional and same pairs
of $\braket{N}$ and $\braket{\delta N^2}$ are always chosen.
However in multi-field cases, 
the trajectory in the field space is diverse and different pairs
of $\braket{N}$ and $\braket{\delta N^2}$ can be selected depending on the trajectory. 
Remembering that the power spectrum is obtained from 
$\braket{\delta N^2}(\braket{N})$ (see eq.~\eqref{Pzeta calculation}),
one can see that $\mathcal{P}_\zeta$ depends on the  trajectory.
Furthermore, it should be noted that many different trajectories are actually realized depending on the spatial points within our observable universe.
Therefore we should take a statistical average of the various trajectories to obtain $\mathcal{P}_\zeta$.

Because of these issues, to obtain $\braket{\delta N^2}$ as a function of $\braket{N}$,
it is needed to take the statistical average over the solutions of Langevin equation namely the trajectories of inflatons
which are realized in the observable universe.
Since our observable universe was 
in one Hubble patch at about 60 e-folds before the end of inflation, 
the diverse solutions should have the same set of field values at that time.
Specifically, we propose a following procedure.
\begin{itemize}
\item[i.]Set the initial condition corresponding to the time of $\braket{N}\sim60$.

\item[ii.]Solving the Langevin equations numerically from this initial value repeatedly, obtain a lot of solutions $\phi^I(N)$ where the superscript $I$ denotes 
different inflatons.
These solutions are used as trajectories in the inflaton field space and they are called sample paths.

\item[iii.]For one sample path, taking a ``initial" value $\phi^{I}_{i}$ on that path, one power spectrum can be obtained by the algorithm
mentioned above.
Other power spectra can also be obtained for other sample paths and the true power spectrum averaged over our observable universe is 
obtained by averaging these power spectra. 

\end{itemize}

\begin{figure}
        \begin{center}
                \includegraphics[width=8cm]{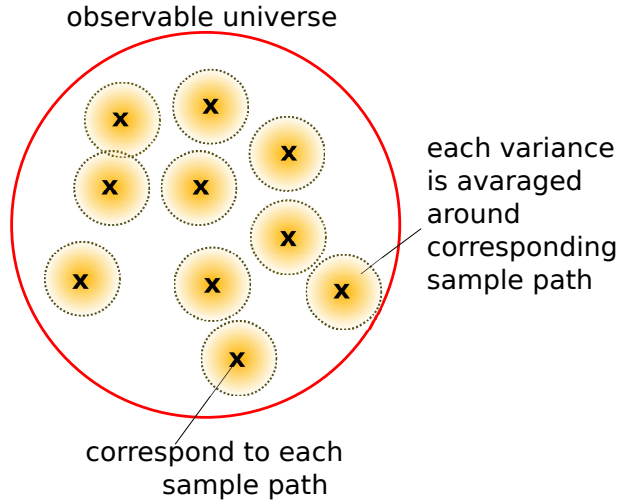}
                \caption{Each sample path corresponds the inflaton dynamics at some spatial point in the observable universe.
                The variance as a function $\braket{N}$ can be obtained from the algorithm 1.-4. for each sample path,
                but these are averaged over only around the corresponding sample path. Therefore, the variance averaged over the observable
                universe is approximated by averaging these variance again.}
                \label{stat_ave}
        \end{center}
\end{figure}

Let us describe why we should take an average over the sample paths in more detail. Since these sample paths branch off at $\braket{N}\sim60$,
their corresponding spatial points are in the same current Hubble patch, namely the observable universe.\footnote{Of course, 
this observable universe is not necessarily \emph{our} observable universe. It corresponds just to a current Hubble scale of
somewhere. However if the inflatons trace some attractor at $\braket{N}\sim60$, every current observable universe should have almost same
structure and the given power spectrum represents also that of our observable universe.}
The variance $\braket{\delta N^2}$ for a sample path is the averaged value over the spatial region just around the spatial point corresponding 
to the sample path, because $\braket{\delta N^2}$
are computed from solutions branching off from the sample path. Therefore the variance averaged over the observable universe is
well approximated by averaging these variances again over many sample paths. See also figure~\ref{stat_ave}.

The power spectrum of super-horizon coarse-grained curvature perturbations can be calculated by the above algorithms 
even in a case where the perturbative expansion with respect to some inflaton field is invalid because the above algorithm does not need a perturbative
expansion with respect to the inflaton field on super-horizon scale.
Since the result depends on the initial condition at $\braket{N}\sim60$, it is required that inflaton dynamics is on some attractor at that time
to make the model predictive, but it is not needed that inflatons always trace an attractor from the beginning to the end of inflation.
In the subsequent two sections, we will apply the stochastic-$\delta N$ to two well-known inflation models, chaotic inflation
and hybrid inflation.

\section{Chaotic inflation}\label{chaotic}
In this section, we will apply the stochastic-$\delta N$ to chaotic inflation as a demonstration.

Chaotic inflation~\cite{Linde:1983gd} is a simplest single-large-field inflation. The potential is given by the mass term $V(\phi)=\frac{1}{2}m^2\phi^2$
or the quartic term $V(\phi)=\frac{\lambda}{4}\phi^4$. More generally, the case where the potential is described as $V(\phi)=\frac{\lambda\phi^n}
{nM_p^{n-4}}$ is also called chaotic inflation. Here we consider the mass term type chaotic inflation model with $V(\phi)=\frac{1}{2}m^2\phi^2$.

To see the dynamics of chaotic inflation, let us calculate the slow-roll parameters, eq.~(\ref{slow-roll parameter}). 
For the mass term potential, these parameters read
\begin{eqnarray}\label{chaotic slow-roll}
        \epsilon_\phi=\eta_{\phi\phi}=\frac{2M_p^2}{\phi^2}.
\end{eqnarray}
Therefore, if the inflaton field has a super-Planckian value $\phi\gg M_p$, inflation takes place. Just about Planck time after the beginning 
of the universe (it may be called ``chaotic" phase), the inflaton can have approximately Planck energy by the quantum effect:
\begin{eqnarray}
        V(\phi)=\frac{1}{2}m^2\phi^2\sim M_p^4.
\end{eqnarray}
Accordingly, if the inflaton mass $m$ is smaller enough than the Planck mass, the inflaton field can get a super-Planckian value naturally as follows.
\begin{eqnarray}
        \phi\sim\left(\frac{M_p}{m}\right)M_p\gg M_p.
\end{eqnarray}
Thus chaotic inflation is free from the initial condition problem. When the inflaton rolls down below $\phi_f=\sqrt{2}M_p$, 
the slow-roll parameters~(\ref{chaotic slow-roll}) exceed unity and inflation ends. The solution of the slow-roll eq.~(\ref{slow-roll eq})
is given by
\begin{eqnarray}
        \phi(N)=\sqrt{4N+2}M_p.
\end{eqnarray}
Here $N$ denotes the e-folds taken from $\phi(N)$ to $\phi_f=\sqrt{2}M_p$.

\begin{figure}
        \begin{center}
                \begin{tabular}{cc}
                        \begin{minipage}{6.5cm}
                                \includegraphics[width=6cm]{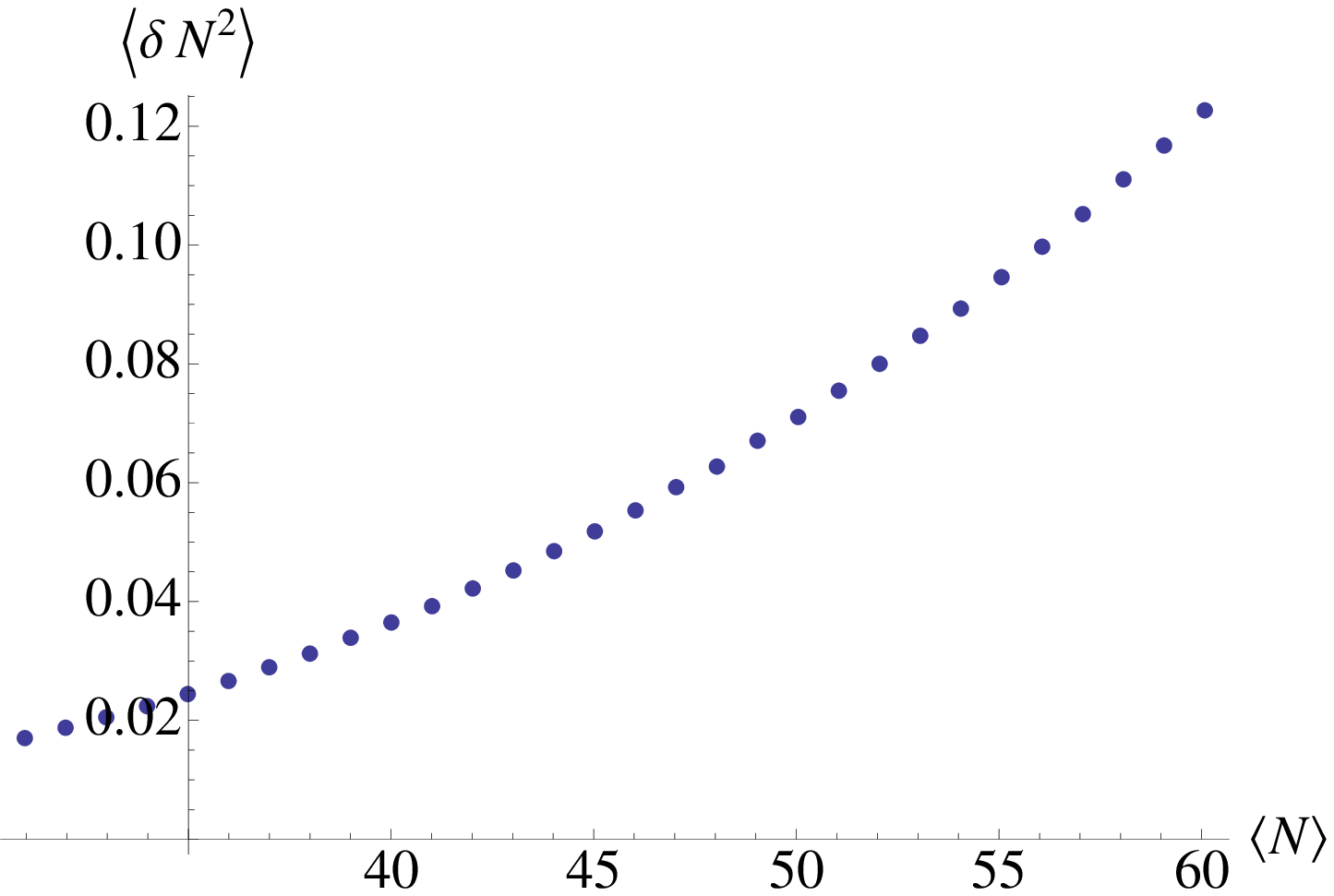}
                        \end{minipage}
                        \begin{minipage}{6.5cm}
                                \includegraphics[width=6cm]{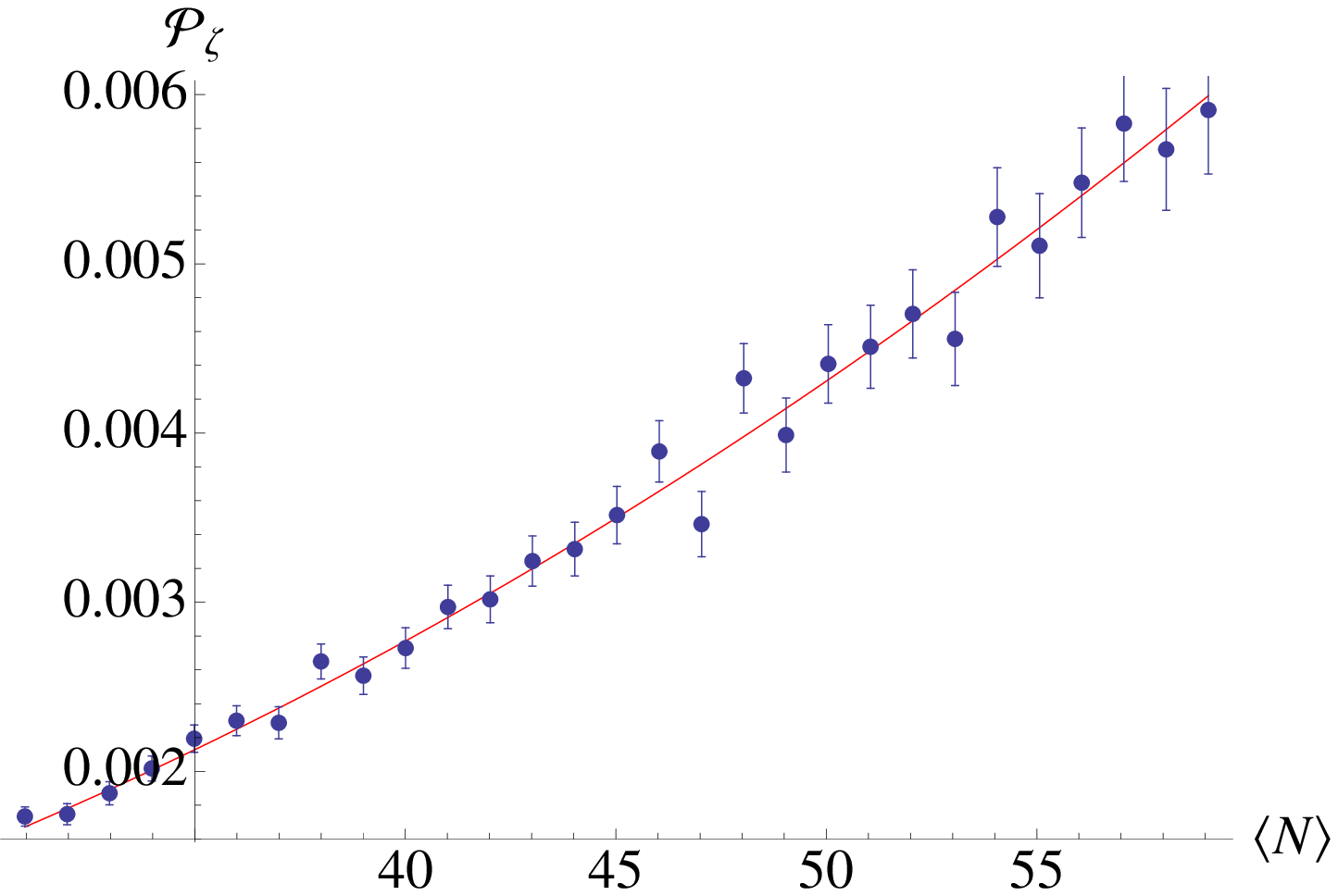}
                        \end{minipage}
                \end{tabular}
                \caption{The relation between the mean taken e-folds $\braket{N}$ and the variance of those $\braket{\delta N^2}$ (left panel)
                and the power spectrum which is a derivative of the plot in the left panel with the inflaton mass $m=0.01M_p$ (right panel). In the 
                plot in the left panel,
                one point corresponds to one ``initial" value $\phi_i$ respectively, and we make 800,000 realizations for each 
                ``initial" value. In the right panel, the red line represents the result of the standard linear perturbation theory~(\ref{linear result}).
                It can be read from this figure that the stochastic-$\delta N$ is quite consistent with the linear perturbations theory in single-field
                inflation as we proved in the previous paper~\cite{Fujita:2013cna}. Note that we evaluate the error for both the variance
                and the power spectrum but that for the variance is so small that the error bars cannot be seen.}
                \label{chaoticdata}
        \end{center}
\end{figure}

Then let us calculate the power spectrum of the curvature perturbations with use of the stochastic-$\delta N$. Since 
chaotic inflation is a single-field model, there is no difficulty and what to do is just to obtain the variance of e-folds $\braket{\delta N^2}$
as a function of the mean of e-folds $\braket{N}$.
In the left panel of figure~\ref{chaoticdata}, we show the relation between $\braket{N}$ and $\braket{\delta N^2}$ 
with the inflaton mass $m=0.01M_p$.\footnote{To be consistent with the CMB observation, 
the inflaton mass should be about $10^{13}\,\mathrm{GeV}$. However, since
we want to show only how to use the stochastic-$\delta N$ algorithm and it is not our goal to construct the viable inflationary model.
Therefore we choose a rather large inflaton mass to make the curvature perturbations larger and $\braket{\delta N^2}$ more conspicuous.}
Different points of this plot correspond to different ``initial" values $\phi_i$. For example, if one choose the ``initial" value
$\phi_i=\phi(N=50)=\sqrt{4\times50+2}M_p$, the mean e-folds $\braket{N}$ will be approximately 50, and at that time,
 it can be read from figure~\ref{chaoticdata} that the variance of e-folds will be about $0.07$. In this paper, we make 800,000 realizations
for each point.

Differentiating the plot in the left panel, we can obtain the power spectrum shown in the right panel of figure~\ref{chaoticdata}.
The red line represents the result of the standard linear perturbation theory~(\ref{linear result}).
As we showed in ref.~\cite{Fujita:2013cna}, the result of the stochastic-$\delta N$ is quite consistent with that of the linear perturbation theory
in single-field inflation. That is because, in single-field inflation, the second and higher-order terms in the perturbative expansion of 
$\zeta$ 
are suppressed by the slow-roll parameters and then the linear approximation of $\zeta$ is good enough.

Note that the errors to the power spectrum are relatively large, even though those to the variance are so small that
the error bars cannot be seen in figure~\ref{chaoticdata}.
In the stochastic-$\delta N$, we do not directly calculate the power spectrum but obtain the variance first, then 
the errors $\Delta\braket{\delta N^2}$ 
are proportional to the variance, $\Delta\braket{\delta N^2}\propto\braket{\delta N^2}$.\footnote{For example, 
here we suppose that $N$ follows some distribution whose true variance is $\sigma^2$.
From the data of e-folds $\{N_i\},i=1,2,\cdots,n$, we can obtain the sampling variance $\braket{\delta N^2}=\frac{1}{n}\sum_{i=1}^n(N_i-\braket{N})^2$,
and the variance of the sampling variance is given by 
$\mathbb{E}\left[(\braket{\delta N^2}-\sigma^2)^2\right]$. Here $\mathbb{E}$ denotes the expected value under the assumed distribution. 
By straightforward but tedious calculations, it is shown that this value is approximated by 
$\frac{1}{n^2}\sum_{i=1}^n(N_i-\braket{N})^4-\frac{\braket{\delta N^2}^2}{n}$ for large $n$. Assuming the Gaussian distribution, 
$\frac{1}{n}\sum_{i=1}^n(N_i-\braket{N})^4\simeq3\sigma^4$ and therefore the error of $\braket{\delta N^2}$ is given by
\begin{eqnarray}
        \left(\mathbb{E}\left[(\braket{\delta N^2}-\sigma^2)^2\right]\right)^{1/2}\simeq\sqrt{\frac{3\sigma^4}{n}-\frac{\sigma^4}{n}}
        =\sqrt{\frac{2}{n}}\sigma^2\simeq\sqrt{\frac{2}{n}}\braket{\delta N^2}.
\end{eqnarray}
In this section, we use this error.
}
Since the power spectrum is connected to the variance by differentiation, which is a linear 
operator, the errors are propagated linearly and those of the power spectrum are also proportional to the variance.
Indeed the power spectrum which is obtained by the finite difference of the variance is given by
\begin{eqnarray}
        \mathcal{P}_\zeta(k=k_fe^{-\braket{N}_i})&=&\frac{(\braket{\delta N^2}_{i+1}\pm\Delta\braket{\delta N^2}_{i+1})
        -(\braket{\delta N^2}_i\pm\Delta\braket{\delta N^2}_i)}
        {\braket{N}_{i+1}-\braket{N}_i} \nonumber \\
        &=&\braket{\delta N^2}_{i+1}-\braket{\delta N^2}_i\pm\left(\Delta\braket{\delta N^2}_{i+1}+\Delta\braket{\delta N^2}_i\right),
\end{eqnarray}
where $\Delta\braket{\delta N^2}_i$ denotes the error of $\braket{\delta N^2}_i$ and we set $\braket{N}_{i+1}-\braket{N}_i=1$.
Therefore the error of the power spectrum is $\Delta\braket{\delta N^2}_{i+1}+\Delta\braket{\delta N^2}_i\simeq2\Delta\braket{\delta N^2}_i$.
On the other hand, 
if the power spectrum is a nearly scale-invariant, the variance can be approximated by $\braket{\delta N^2}\sim\braket{N}\mathcal{P}_\zeta$
from eq.~(\ref{variance and power}).
Hence the errors of the power spectrum are relatively sizable for a large $\braket{N}$. Because of this fact, the stochastic-$\delta N$
approach is not so adequate to calculate the large-scale power spectrum.
In contrast, to calculate the small-scale power spectrum, it is quite useful. 
As we will see in the next section, the stochastic-$\delta N$ approach enables the calculation of the large peak profile on small scales.

\section{Hybrid inflation}\label{hybrid}
\subsection{Overview of the original type}
Hybrid inflation~\cite{Linde:1993cn,Copeland:1994vg} is an intriguing inflation model combining chaotic inflation and new inflation.
This model does not need super-Planckian field value likes as new inflation, and moreover, the initial condition problem is softened than 
new inflation in a similar way to chaotic inflation.
The extensions to supersymmetric (SUSY) types are also studied well~\cite{Halyo:1996pp,Binetruy:1996xj,Dvali:1994ms,Kallosh:2003ux}.

In hybrid inflation, there exist two scalar fields, one is an inflaton $\phi$ and the other is a
waterfall field $\psi$.
The potential of the original type is given by
\begin{eqnarray}\label{hybrid potential}
        V(\phi,\psi)&=&\Lambda^4\left[\left(1-\frac{\psi^2}{M^2}\right)^2+\frac{\phi^2}{\mu^2}+2\frac{\phi^2\psi^2}{\phi_c^2M^2}\right] \nonumber \\
        &=&\Lambda^4+\frac{1}{2}\left(\frac{2\Lambda^4}{\mu^2}\right)\phi^2+\frac{1}{2}\left[\frac{4\Lambda^4}{M^2}\left(\frac{\phi^2}{\phi_c^2}-1\right)
        \right]\psi^2+\frac{\Lambda^4}{M^4}\psi^4,
\end{eqnarray}
with the model parameters $\Lambda,\mu,M$ and $\phi_c$. The dynamics of this inflation is as follows. For an appropriate initial condition,
the $\psi$ field settles down to $\psi=0$ due to the term of $\psi^4$. Then 
inflation is driven by the constant potential $V_0=\Lambda^4$, $\phi$ rolling down slowly due to the inflaton mass $m_\phi^2=2\Lambda^4/\mu^2$.
The key point is the waterfall mass $m_\psi^2=\frac{4\Lambda^4}{M^2}\left(\frac{\phi^2}{\phi_c^2}-1\right)$ becomes negative when $\phi$ rolls down
below $\phi_c$. Then, the waterfall field rolls down rapidly to the potential minimum $(\phi,\psi)=(0,\pm M)$ and slow-roll inflation is over.
The point $(\phi,\psi)=(\phi_c,0)$ is called ``critical point", and the phases before and after the critical point
are called ``valley phase" and ``waterfall phase", respectively. Even though  the waterfall phase usually ends rapidly, it
is possible that the waterfall phase lasts for more than 10 e-folds with some parameters. We consider such a case in this paper,
since we are interested in a peak profile in power spectrum because of the flatness of the potential around the critical point. 

The power spectrum of the curvature perturbations during the valley phase can be calculated easily because in this phase, 
$\psi$ settles down to zero and it is almost single-field inflation. At this epoch, the slow-roll parameter 
$\epsilon_\phi$~(\ref{slow-roll parameter}) reads
\begin{eqnarray}
        \epsilon_\phi=\frac{M_p^2}{2}\left(\frac{m_\phi^2\phi}{\frac{1}{2}m_\phi^2\phi^2+V_0}\right)^2
        \simeq\frac{M_p^2m_\phi^4\phi^2}{2V_0^2}.
\end{eqnarray}
Therefore, the power spectrum~(\ref{linear result}) is written as
\begin{eqnarray}\label{power during valley}
        \mathcal{P}_\zeta(k)\simeq\left.\frac{1}{12\pi^2M_p^6}\frac{V_0^3}{m_\phi^4\phi^2}\right|_{k=aH}.
\end{eqnarray}

On the other hand, the power spectrum around the critical point and during
the waterfall phase is not solved fully analytically. 
After $\phi$ approaches the critical point, not only $\phi$ but also $\psi$ contribute the inflation dynamics and
the curvature perturbations. 
Around the critical point $(\phi_c,0)$, the potential is extremely flat 
in the direction of $\psi$ as
easily checked from eq.~\eqref{hybrid potential}. Therefore 
the quantum fluctuations of $\psi$ surpass the zero mode, namely $\braket{\delta\psi^2}\gg\psi_0^2$, and then the perturbations with respect to $\psi$
are broken down. 
Many authors calculated the power spectrum during the waterfall phase for special cases and 
Lyth provided more general treatment when the linear approximation
of e.o.m. is good~\cite{Lyth:2010zq,Lyth:2012yp}. 
This paper gives the full solution using the stochastic formalism.
Naively speaking, because of the flat potential, the curvature perturbations will rapidly grow and show peak profile.
In subsection~\ref{dynamics and power spectrum}, we will see the calculated curvature perturbations indeed show such a peak profile.

\subsection{Amplitude of noise}
Before calculate the power spectrum, we should mention the amplitude of the noise term. As showed in section \ref{stochastic delta N},
the noise term is proportional to the power spectra of the scalars evaluated at the horizon crossing $k=\epsilon aH$.
Then let us consider the evolution of the sub-horizon mode. Similarly to eq.~(\ref{lin eom}), we linearize the e.o.m. with respect to $\phi_\mathbf{k}$
as
\begin{eqnarray}\label{linearized eom}
        \ddot{\phi}_{\mathbf{k}}^I+3H(\phi_\mathrm{IR},\psi_\mathrm{IR})\dot{\phi}_{\mathbf{k}}^I+
        \left(\frac{k}{a}\right)^2\phi_{\mathbf{k}}^I+V_{IJ}(\phi_\mathrm{IR},\psi_\mathrm{IR})
        \phi_{\mathbf{k}}^J =0,
\end{eqnarray}
where superscript $I,J$ denote $\phi$ and $\psi$. This linearization requires $\phi_{\mathrm{IR}}^I\gg\phi_{\mathrm{UV}}^I$ where
$\phi_{\mathrm{UV}}^I$ is a sum of the sub-horizon modes $k>\epsilon aH$, namely
$\phi_{\mathrm{UV}}^I(t,\mathbf{x})=\int\frac{d^3k}{(2\pi)^3}\theta(k-\epsilon aH)\phi_\mathbf{k}^I(t)e^{-i\mathbf{k}\cdot\mathbf{x}}$. 
Note that $\phi_\mathrm{IR}^I(t,\mathbf{x})+\phi_\mathrm{UV}^I(t,\mathbf{x})=\phi^I(t,\mathbf{x})$.
If the homogenous zero mode is
much larger the fluctuations of the inflaton, $\phi_{0}^I\gg\delta\phi^I$, as usual, the linearization is valid because the IR part includes
the zero mode. Moreover, even if the zero mode is smaller than the fluctuation, 
the IR part is generally larger than the UV part because the IR part receives the white noise and has the field value of about
the Hubble parameter at least. Therefore, the above linearization is valid in many cases.\footnote{Actually, in our calculation, $\psi_\mathrm{IR}$
remains zero for a little while because we will neglect the noise of $\psi$ when $\psi$ is highly massive as we will mention.
Therefore, the IR part of $\psi$ is smaller than the UV part and 
the linearization of the e.o.m. does not seem to be valid. However, since we will consider the case where the field value of 
$\phi_\mathrm{IR}$ is much larger than the Hubble parameter
(see~(\ref{parameter})), the higher order
term in $V_\psi$, $\psi_\mathrm{UV}^3$, is negligible compared to the $\phi_\mathrm{IR}^2\psi_\mathrm{UV}$ term. 
The other higher order term $\frac{\Lambda^4}{M^2}\frac{\psi_\mathrm{UV}^2}{\phi_c^2}\phi_\mathrm{IR}$
in $V_\phi$ is also negligible compared to $\frac{\Lambda^4}{\mu^2}\phi_\mathrm{IR}$ term.}

Note that the Hubble parameter and the derivatives of the potential 
are the functions of the IR fields $\phi_\mathrm{IR}$ and $\psi_\mathrm{IR}$ evaluated around the spatial point considered here.
However, $\phi_\mathrm{IR}$ and $\psi_\mathrm{IR}$ themselves depend
on the past amplitude of the noise term and thus it is hard to solve the eqs.~(\ref{linearized eom}) in an exact manner. In this paper, we approximate
$\mathcal{P}_\phi$ and $\mathcal{P}_\psi$ by the solution with the constant Hubble parameter and the constant scalar mass, 
eq.~(\ref{const mass power}),
\begin{eqnarray}\label{const mass approx}
        \mathcal{P}_{\phi^I}(k=\epsilon aH)=\frac{H^2}{8\pi}\epsilon^3\left|H_{\nu_I}^{(1)}(\epsilon)\right|^2,
\end{eqnarray}
where the Hubble parameter
and the scalar mass $m_I^2=V_{II}$ are evaluated at the horizon crossing, $k=\epsilon aH$. 
When $m_I^2/H^2$ exceeds $9/4$ and $\nu_I=\sqrt{9/4-m_I^2/H^2}$ becomes
imaginary, the corresponding noise terms will be suppressed by $\epsilon^3$ and negligible, so we omit them in the numerical calculation.

It should also be noted that the different kinds of the inflaton fields interact with each other through the mixing term $V_{IJ}$ 
and then the noise terms for different scalar fields can have non-zero correlations. However, with the parameters considered in this paper,
the effective masses of the inflaton and the waterfall field are quite different, so we can neglect the effect of mixings.\footnote{To 
take the effect of mixings into account, see appendix~\ref{mixing}.}

\subsection{Dynamics and power spectrum}\label{dynamics and power spectrum}
Let us then move to the calculation of the power spectrum.
In this paper, we consider the original type of hybrid inflation whose potential is represented as eq.~(\ref{hybrid potential})
with following parameter values,
\begin{eqnarray}\label{parameter}
        \frac{\mu}{M_p}=100, \quad \frac{M}{M_p}=\frac{\phi_c}{M_p}=0.13, \quad \frac{\Lambda}{M_p} = 1.2\times10^{-4},
\end{eqnarray}
With an above value, it takes about 15 e-folds from the critical point to the end of inflation. The energy scale $\Lambda$ is
determined so that the amplitude of the curvature perturbations during a valley phase satisfies 
the observed value ($\mathcal{P}_\zeta^{1/2}\sim5\times
10^{-5}$)~\cite{Ade:2013zuv}.\footnote{Since our goal is not to construct an inflationary model as mentioned in the footnote of the previous section,
there is no need to set $\mathcal{P}_\zeta\sim(5\times10^{-5})^2$ actually.}

\begin{figure}
        \begin{center}
                \begin{tabular}{cc}
                        \begin{minipage}{6.5cm}
                                \includegraphics[width=6cm]{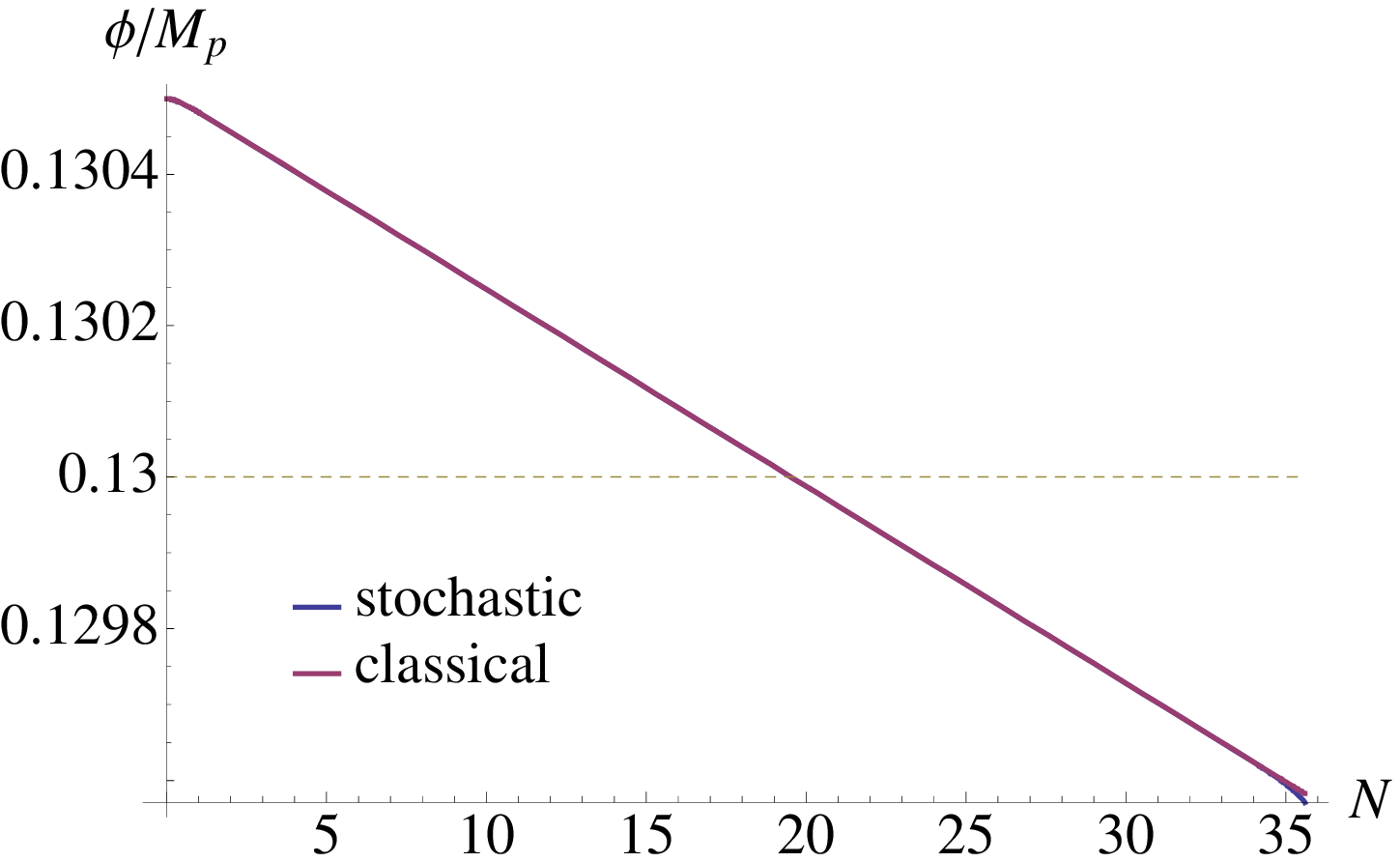}
                        \end{minipage}
                        \begin{minipage}{6.5cm}
                                \includegraphics[width=6cm]{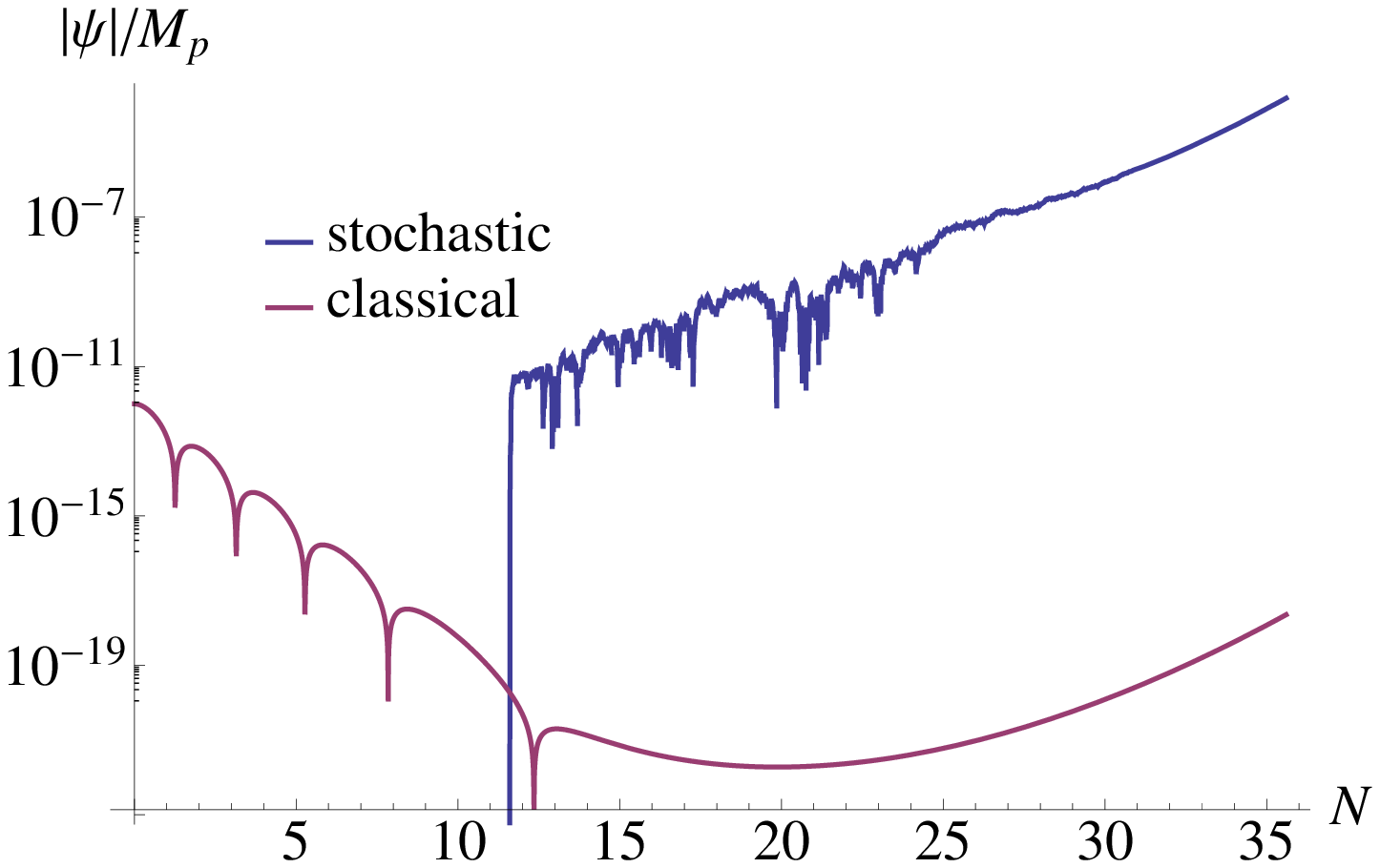}
                        \end{minipage}
                \end{tabular}
                \caption{The plot of one sample path with white noise (blue line) and a classical path without noise (red line). 
                The initial condition for the stochastic path is 
                $(\phi_i/M_p,\psi_i/M_p)=(0.1305,0)$ and that for the classical path is $(\phi_i/M_p,\psi_i/M_p)=(0.1305,10^{-12})$. 
                We set the non-zero initial value of $\psi$ for the classical path because otherwise it remains zero without noise.
                The inflaton $\phi$ is not significantly affected by quantum noise  but
                the waterfall $\psi$ grows rapidly due to the noise after the mass of $\psi$ becomes less than the Hubble parameter.
                Inflation without noise
                continues about 55 e-folds, while the stochastic cases end faster by about 20 e-folds than the classical case 
                because of the rapid growth of $\psi$. In both cases,
                the inflaton field reaches the critical point $\phi_c=0.13M_p$ (dotted line in left panel) about 20 e-folds after the beginning of inflation.}
                \label{drefpath}
        \end{center}
\end{figure}

Since the $q_\nu$ terms in the Langevin equation are neglected,
the e.o.m. is written as
\begin{eqnarray}
        \begin{cases}
                \displaystyle
                \dif{\phi}{N}(N)=\frac{\pi_{\phi}}{H}(N)+\mathcal{P}_{\phi}^{1/2}(N)\xi_{\phi}(N), \\[5pt]
                \displaystyle
                \dif{\pi_{\phi}}{N}(N)=-3\pi_{\phi}(N)-V_{\phi}, \\[5pt]
                \displaystyle
                \dif{\psi}{N}(N)=\frac{\pi_\psi}{H}(N)+\mathcal{P}_\psi^{1/2}(N)\xi_\psi(N), \\[5pt]
                \displaystyle
                \dif{\pi_\psi}{N}(N)=-3\pi_\psi(N)-V_\psi,
        \end{cases}
\end{eqnarray}
where $\mathcal{P}_{\phi(,\psi)}(N)$ is approximated by $\frac{H^2}{8\pi}\epsilon^3\left|H^{(1)}_{\nu_{\phi(,\psi)}}(\epsilon)\right|^2$ with
$\nu_{\phi(,\psi)}=\sqrt{\frac{9}{4}-\frac{V_{\phi\phi(,\psi\psi)}}{H^2}}$ and $\xi_\phi$ and $\xi_\psi$ are zero-mean independent white noises:
\begin{eqnarray}
        \braket{\xi_\phi(N)\xi_\phi(N^\prime)}&=&\braket{\xi_\psi(N)\xi_\psi(N^\prime)}=\delta(N-N^\prime), \nonumber \\
        \braket{\xi_\phi(N)\xi_\psi(N^\prime)}&=&0.
\end{eqnarray}
Note that we use the dimensionless e-folds $dN=Hdt$ as a time variable instead of the cosmic time $t$. $\xi_R(t)$ and $\xi_{\phi(,\psi)}(N)$ are 
connected by change of variables of the delta function:
\begin{eqnarray}
        \braket{\xi_R(t)\xi_R(t^\prime)}=\delta(t-t^\prime)=H\delta(N-N^\prime)=H\braket{\xi_{\phi(,\psi)}(N)\xi_{\phi(,\psi)}(N^\prime)}.
\end{eqnarray}

The time evolutions of the inflaton and waterfall field on one sample path with  
the initial condition $(\phi_i/M_p,\psi_i/M_p)=(0.1305,0)$ 
are shown in figure~\ref{drefpath} as ``stochastic". For the sake of comparison, we also plot the solution without noise as ``classical",
with tiny but non-zero initial $\psi$ because the $\psi$ field with $\psi_i=0$ remains zero thereafter without noise.
$N=0$ corresponds to the beginning of inflation. The waterfall field $\psi$ remains zero at first, because
the mass of $\psi$ is as large as the Hubble parameter (or, $\nu_\psi$ is imaginary) and we omit the quantum noise as mentioned in 
the previous
subsection. Subsequently, due to the noise, 
$\psi$ grows much rapidly
compared with the classical solution. The inflaton field $\phi$ seems not to be affected by the noise term 
and its dynamics is almost same in both the stochastic and classical case.
However, with the parameters in this paper,
the end of inflation is determined by the value of $\psi$. 
In fact, inflation without noise continues about 20 e-folds longer than stochastic inflation
because the field value of $\psi$ does not grow fast without noise.
It shows the importance of the stochastic effect not only for the calculation of the curvature perturbations but also the background dynamics.
Note that, in both cases, the critical phase $\phi_c=0.13M_p$ is reached about 20 e-folds after the beginning of the calculation.

\begin{figure}
        \begin{center}
                \includegraphics[width=10cm]{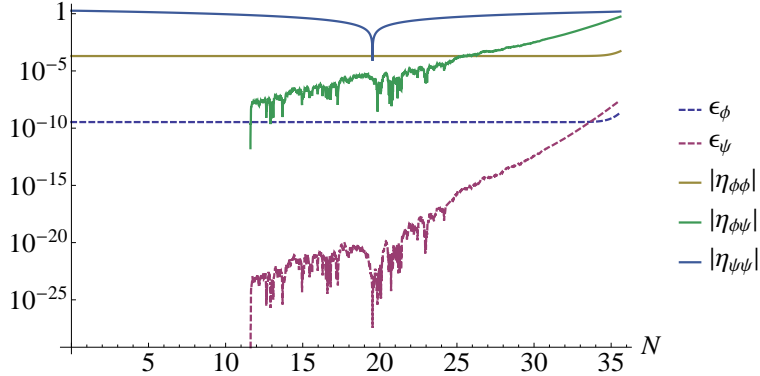}
                \caption{The time dependencies of 5 slow-roll parameters for stochastic inflation. For $N\ltsim20$, 
                since inflation is in the valley phase, $\eta_{\psi\psi}$ is not a relevant slow-roll parameter and it does not matter that 
                $|\eta_{\psi\psi}|>1$. For $N\gtsim20$, $|\eta_{\psi\psi}|$ exceeds unity at $N\sim30$.}
                \label{slow-roll}
        \end{center}
\end{figure}

In addition, let us check the time developments of the slow-roll parameters. In two-scalar inflation, there are following 5 slow-roll parameters.
\begin{eqnarray}
        &\displaystyle
        \epsilon_\phi=\frac{M_p^2}{2}\left(\frac{V_\phi}{V}\right)^2, \quad \epsilon_\psi=\frac{M_p^2}{2}\left(\frac{V_\psi}{V}\right)^2 \nonumber \\
        &\displaystyle
        \eta_{\phi\phi}=M_p^2\frac{V_{\phi\phi}}{V} , \quad \eta_{\phi\psi}=M_p^2\frac{V_{\phi\psi}}{V} , \quad \eta_{\psi\psi}=M_p^2\frac{V_{\psi\psi}}{V}.
\end{eqnarray}
The time developments of these slow-roll parameters for the sample path showed in figure~\ref{drefpath} are illustrated in figure~\ref{slow-roll}.
$|\eta_{\psi\psi}|$ exceeds unity first at $N\sim30$ and then slow-roll inflation ends.

\begin{figure}
        \begin{center}
                \includegraphics[width=10cm]{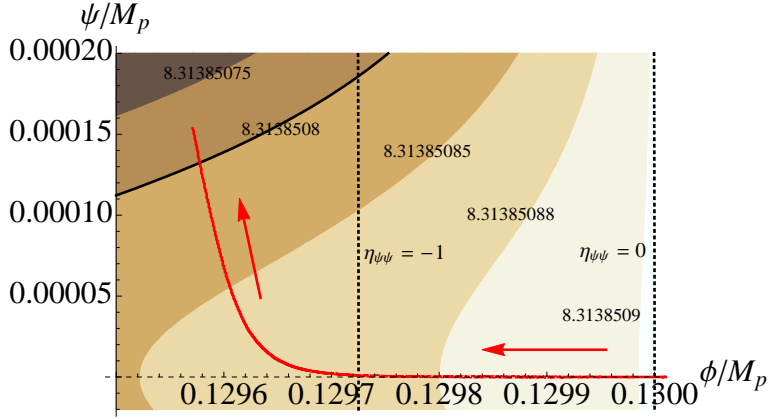}
                \caption{The contour plot of the potential. The red line shows the sample path displayed figure~\ref{drefpath}
                and the dashed lines represent uniform $\eta_{\psi\psi}$ lines respectively. Parameters on the contour lines are the values
                of $\sqrt{V/3M_p^4}\times10^9\simeq (H/M_p)\times10^9$. In this paper, we use $H=8.3138508\times10^{-9}M_p$ (black thick line) 
                as the end condition of inflation.}
                \label{contour}
        \end{center}
\end{figure}

In multi-field cases, the point where the slow-roll condition is violated is unsuitable for
the end point of the field in the $\delta N$ formula as we mentioned in the footnote
of the previous section. That is because in multi-field inflation, slow-roll violating points are not on an equipotential line, though the end slice 
of the $\delta N$ formula 
should be a uniform density slice. Therefore we use the uniform Hubble slice as the end. In figure~\ref{contour}, we show the
contour plot of the potential with the sample path shown in figure~\ref{drefpath} (red line) and uniform $\eta_{\psi\psi}$ lines (dashed lines).
It shows that equipotential lines do not correspond to uniform $\eta_{\psi\psi}$ lines indeed. Parameters on the contour lines represent
the value of $\sqrt{V/3M_p^4}\times10^9\simeq(H/M_p)\times10^9$.
In this paper, we use $H=8.3138508\times10^{-9}M_p$ (black thick line) 
as the end slice of $\delta N$ formula where the slow-roll condition is violated enough as 
can be seen in figure~\ref{contour}.

Since above discussions are just for one sample path, one may doubt it depends on the sample paths. However, we checked that realizations
which quite deviate from this path and violate the above discussion almost never occur. Therefore it is valid to use 
$H=8.3138508\times10^{-9}M_p$ as the end slice.\\

\begin{figure}
        \begin{center}
                \begin{tabular}{cc}
                        \begin{minipage}{7.5cm}
                                \includegraphics[width=7cm]{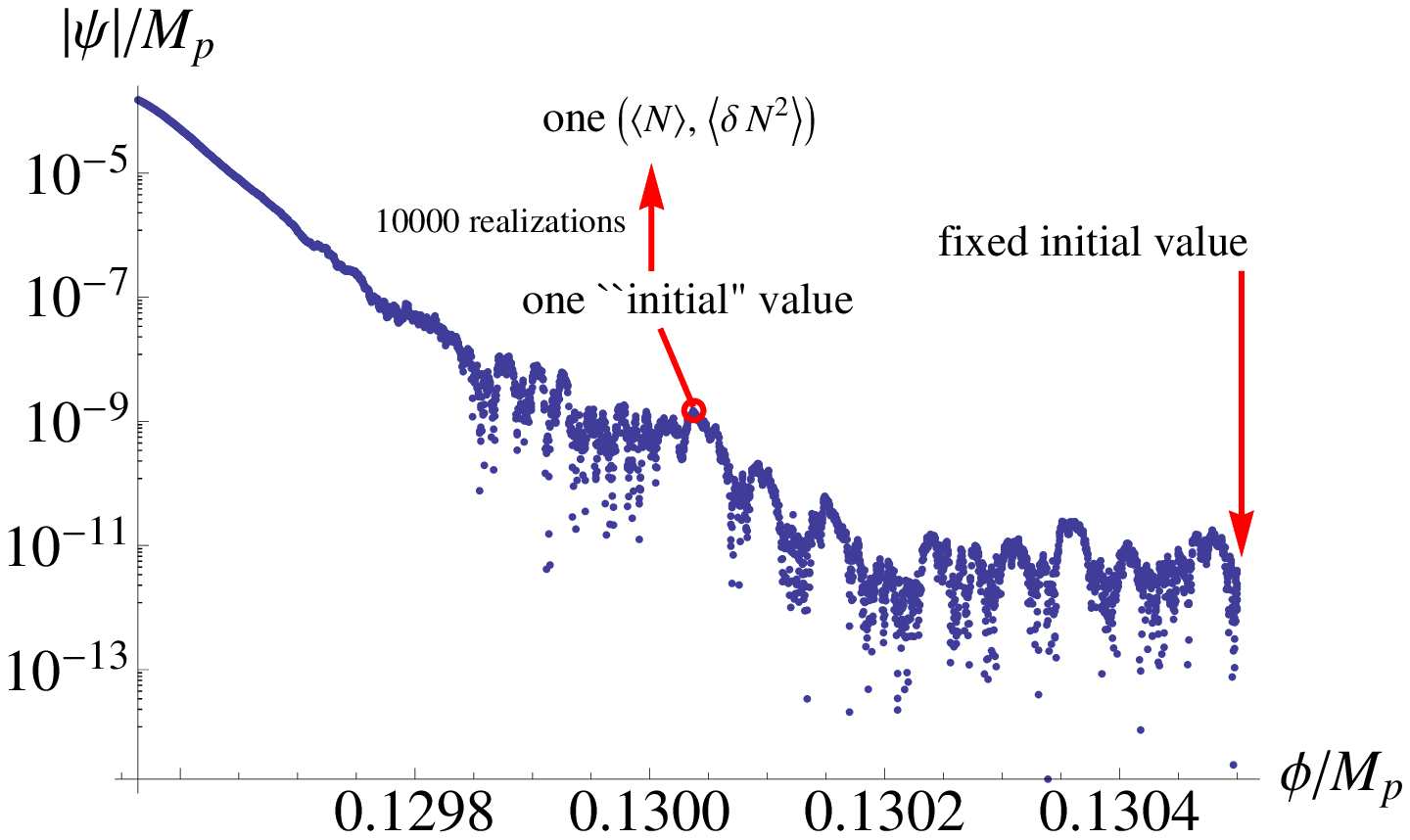}
                        \end{minipage}
                        \begin{minipage}{6.5cm}
                                \includegraphics[width=6cm]{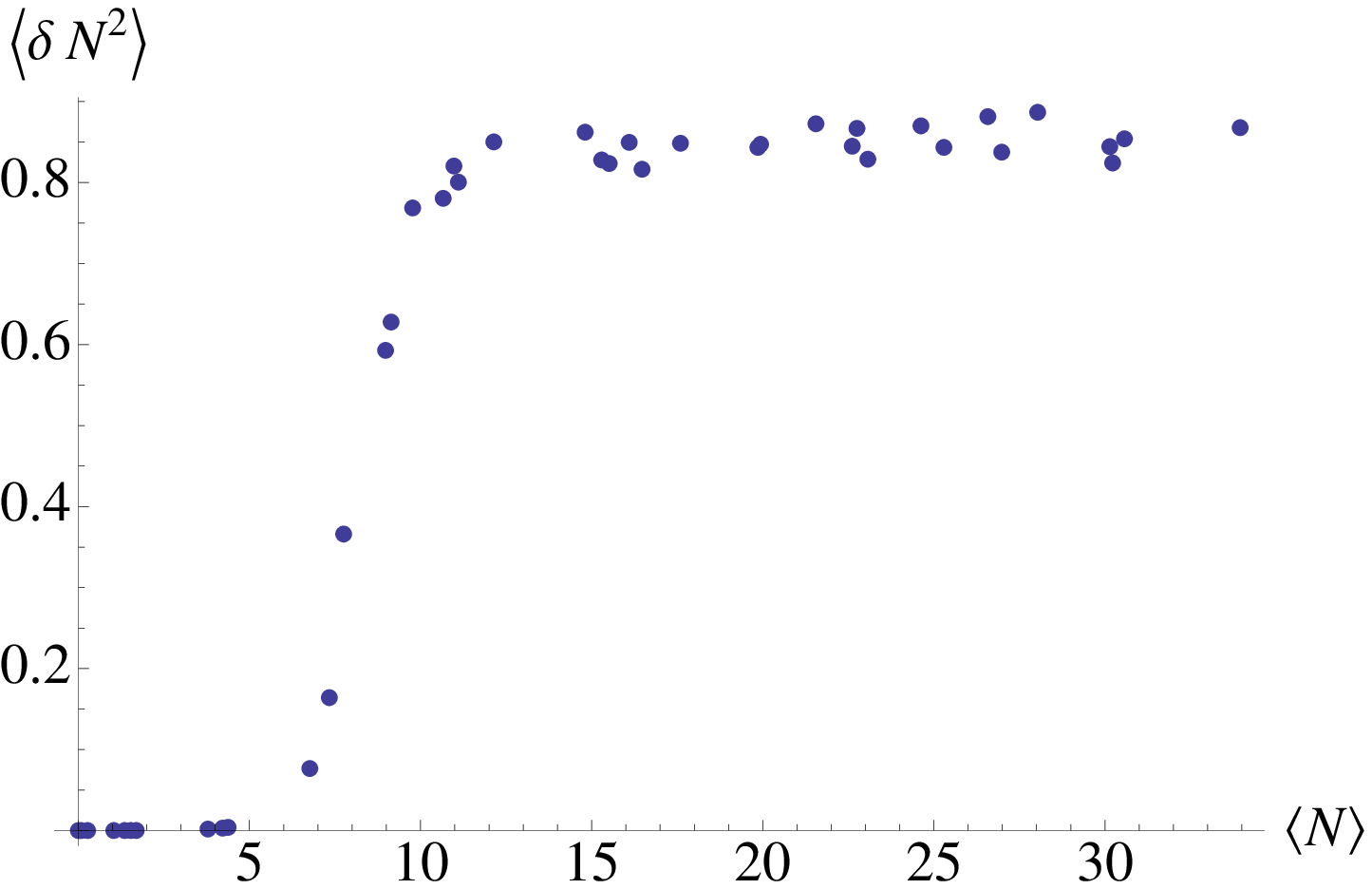}
                        \end{minipage}
                \end{tabular}
                \caption{One sample path started from the fixed initial value $(\phi_i/M_p,\psi_i/M_p)=(0.1305,0)$ (left panel) 
                and corresponding $\braket{N}$ vs. $\braket{\delta N^2}$ plot (right panel). 
                One point on the plot in the right panel corresponds to one ``initial" value
                on that sample path. For each ``initial" value, we make 10000 realizations from that value to the end of inflation and take the average
                and variance of their e-folds.}
                \label{var1}
        \end{center}
\end{figure}

Let us calculate the power spectrum of the curvature perturbations. Recalling the algorithm mentioned 
in section~\ref{stochastic delta N}, we should make many sample paths from the fixed initial condition, 
which we have set at $(\phi_i/M_p,\psi_i/M_p)=(0.1305,0)$.\footnote{This initial value corresponds to $\braket{N}\sim35$ as can be read from figure~\ref{drefpath}, though the initial 
condition should be set on the point which
corresponds to our observable universe $\braket{N}\sim60$ as we mentioned in section~\ref{stochastic delta N}. 
However, in the valley phase, the mass of $\psi$
is heavy enough and $\psi$ rapidly converges to zero. Namely, the point $(\phi_i/M_p,\psi_i/M_p)=(0.1305,0)$ is on an attractor.
Therefore, without specifying the initial condition at $\braket{N}\sim60$, many sample paths converge to around this point.
We can also rephrase it that we chose some initial condition at $\braket{N}\sim60$ from which the sample paths converge to around the point 
$(\phi_i/M_p,\psi_i/M_p)=(0.1305,0)$. In this paper, for simplicity, and since
we are interested only in around the critical point corresponding to $\braket{N}\sim15$, we calculate only about last 35 e-folds.}
For each sample path, we obtain a $\braket{N}$ vs. $\braket{\delta N^2}$ plot taking ``initial" values on that sample path.
For example, we show $\braket{N}$ vs. $\braket{\delta N^2}$ plot for one sample path in figure~\ref{var1}.
Reiterating to solve the Langevin e.o.m. from one ``initial" value on that sample path to the end of inflation, we can get the
e-folds $\braket{N}$ and the variance of them $\braket{\delta N^2}$. 
Then, changing the ``initial" value variously, full $\braket{N}$ vs. $\braket{\delta N^2}$ plot such as the 
right panel of 
figure~\ref{var1} can be obtained. In this paper, we reiterate the calculation 10000 times for each ``initial" value.  

\begin{figure}
        \begin{center}
                \begin{tabular}{cc}
                        \begin{minipage}{6.5cm}
                                \includegraphics[width=6cm]{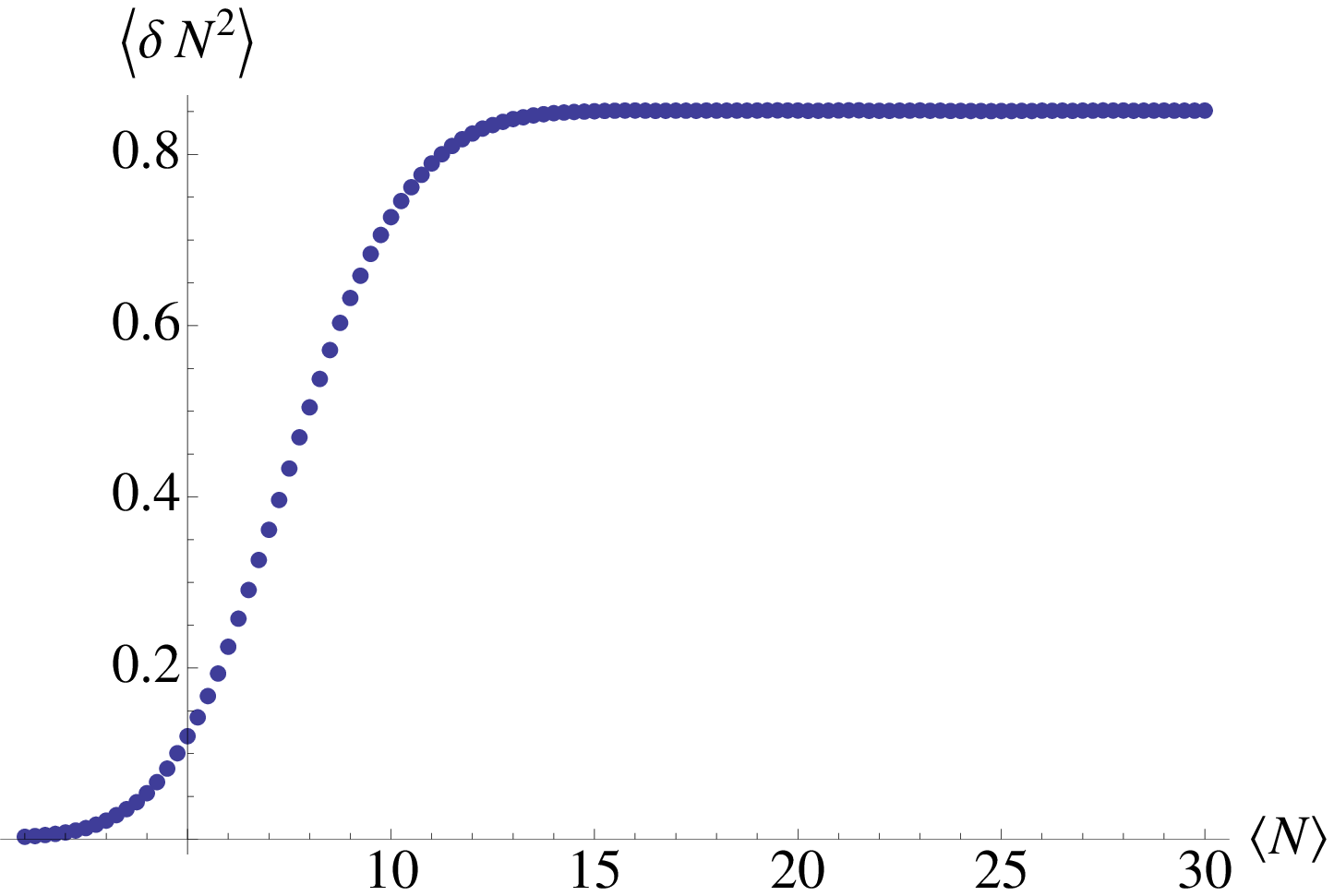}
                        \end{minipage}
                        \begin{minipage}{6.5cm}
                                \includegraphics[width=6cm]{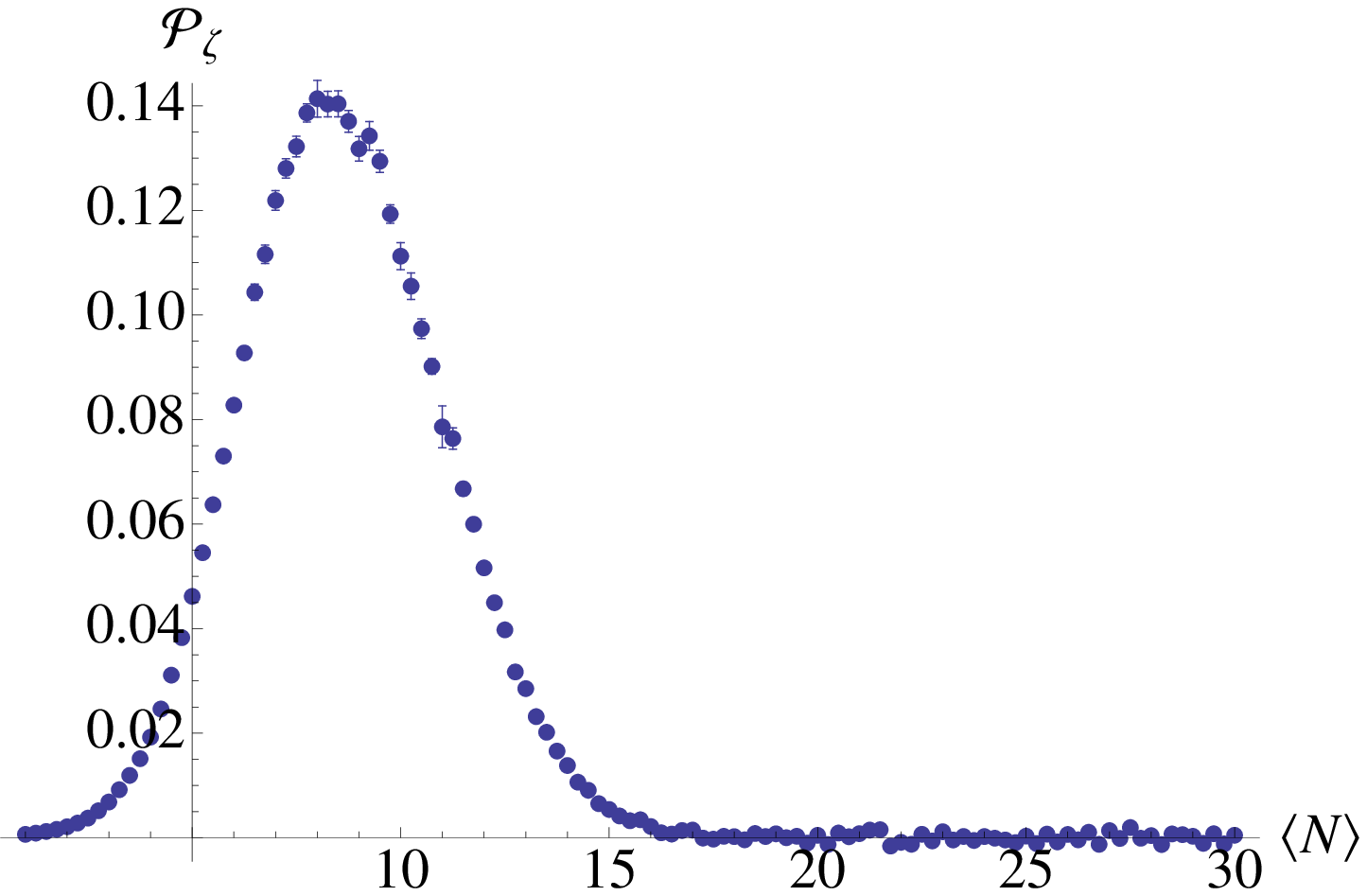}
                        \end{minipage}
                \end{tabular}
                \caption{The $\braket{N}$ vs. $\braket{\delta N^2}$ plot averaged for 10000 sample paths (left panel) and the power spectrum of the curvature
                perturbations as the derivative of that plot (right panel). Error bars represent standard errors. $\braket{N}$ corresponds to a wavenumber by
                the eq. $\braket{N}=\ln(k_f/k)$ where $k_f$ denotes the horizon scale at the end of inflation, $k_f=\epsilon aH|_f$.}
                \label{tpower}
        \end{center}
\end{figure}

Similarly, we can obtain many $\braket{N}$ vs. $\braket{\delta N^2}$ plots for various sample paths. Then the true $\braket{N}$ vs. $\braket{\delta N^2}$
plot of our observable universe is just the average of these plots. In the left panel of 
figure~\ref{tpower}, we show 
the average of $\braket{N}$ vs. $\braket{\delta N^2}$ plots for 10000 sample paths. Finally, differentiating this plot, we obtain the power spectrum
of the curvature perturbations $\mathcal{P}_\zeta=d\braket{\delta N^2}/d\braket{N}$ as shown in the right panel of figure~\ref{tpower}.
The horizontal axis $\braket{N}$ corresponds to a wavenumber $k$ by the relation $\braket{N}=\ln(k_f/k)$
where $k_f$ is the horizon scale at the end of inflation, $k_f=\epsilon aH|_f$, and the scale corresponding to the critical point is $\braket{N}\sim17$
for example. Figure~\ref{tpower} shows the peak of the power spectrum in the waterfall phase after the critical point because of the tachyonic instability
of the waterfall field $\psi$. 

Not only in the hybrid case but also in any highly
stochastic cases, we can calculate the power spectrum applying the stochastic-$\delta N$ formalism shown here. 
Note that for the inflation models or parameters which we adopt, the variance of e-folds 
$\braket{\delta N^2}$ should not exceed unity in our observable universe, namely $\braket{N}\ltsim60$. 
For $\braket{\delta N^2}>1$, the universe becomes too inhomogeneous to account for the observed universe.
Indeed, the constraints from PBHs suggest $\mathcal{P}_\zeta\ltsim10^{-1.5}$ for a wide range of $k$
and therefore $\braket{\delta N^2}=\int\mathcal{P}_\zeta \,d(\log k)\ltsim1$~\cite{Josan:2009qn,Carr:2009jm}.

\section{Conclusion}\label{conclusion}
In this paper, we applied the non-perturbative method that we have proposed in the previous paper~\cite{Fujita:2013cna}, the stochastic-$\delta N$ 
formalism, to chaotic inflation and hybrid inflation. Especially, in hybrid inflation, we chose the parameters 
where the waterfall phase lasts for more than 10 e-folds,
and calculated the power spectrum of the curvature perturbations including around the critical point. The result is shown in figure
\ref{chaoticdata} and \ref{tpower} for chaotic and hybrid inflation respectively. 
In particular, it is the first time that the power spectrum in hybrid inflation is calculated fully from the phase before the
critical point to the end of inflation. The resultant power spectrum shows a peak profile due to 
the tachyonic instability of the waterfall field during the waterfall phase $\braket{N}\ltsim17$. 

Though the recent CMB observations by the Planck and BICEP2 collaborations~\cite{Ade:2013zuv,Ade:2014xna} favor 
a simple single-large-field inflationary model, several multi-field models may be worth considering and they can have the highly stochastic
region. In such cases, the method we demonstrated in this paper is needed to obtain the curvature perturbations.

\acknowledgments

We would like to thank Ryo Namba for helpful discussions and advice.
We also thank the authors of the publicly available distribution ``Mersenne Twister"~\cite{Matsumoto}, 
which we used for generating the stochastic noise in the Langevin equations.
This work is supported by Grant-in-Aid for Scientific research from the Ministry of Education, Science, Sports, and Culture (MEXT), Japan, No. 25400248
[MK], No. 21111006 [MK] and also by World Premier International Research Center Initiative (WPI Initiative), MEXT, Japan.
T.F. acknowledges the support by JSPS Research Fellowships for Young Scientists.
The work of Y.T. is partially supported by an Advanced Leading Graduate Course for Photon Science grant.

\appendix
\section{Numerical calculation of stochastic process}
In this appendix, we comment on the numerical calculation method of the stochastic process. 
There are many numerical integrating methods
with excellent converging properties like a Runge-Kutta method for ordinary differential equations, while the methods for stochastic differential equations
like a Langevin equation are still developing. Since applying the method for ordinary differential equations to stochastic ones directly generally violate
the desired properties of the stochastic process, specific methods should be constructed.
We will give the terminology of stochastic calculus first, and then describe the numerical integrating method we used in this paper.
Note that this appendix is based on ref.~\cite{ShreveII,Kloeden}.

\subsection{Stochastic calculus}
In the first place, let us define \emph{Brownian motion}, which is the simplest stochastic process. 
Brownian motion is the continuous time limit of a random walk and mathematically defined as follows.
\begin{definition}
Some stochastic continuous function $W(t),t\ge0$ is assumed to exist, satisfying $W(0)=0$. Then, if for all $0=t_0<t_1<\cdots<t_m$,
the increments
\begin{eqnarray}
        W(t_1)=W(t_1)-W(t_0),\,W(t_2)-W(t_1),\,\cdots,\,W(t_m)-W(t_{m-1}),
\end{eqnarray}
are independent, Gaussian distributed and satisfy the condition,
\begin{eqnarray}\label{Brownian condition}
        \braket{W(t_{i+1})-W(t_i)}=0, \quad \braket{(W(t_{i+1})-W(t_i))^2}=t_{i+1}-t_i, \quad \text{for all $i$}
\end{eqnarray}
$W(t)$ is called Brownian motion. 
\end{definition}
The zero-mean white noise $\xi(t)$ in the Langevin equation is formally defined as \emph{the derivative of Brownian motion}:\footnote{Strictly speaking, 
Brownian motion is differentiable nowhere. The definition~\ref{def of noise} is just a formal one.}
\begin{eqnarray}\label{def of noise}
        \xi(t)=\dif{W(t)}{t}, \quad \text{or} \quad W(t)=\int^t_0\xi(t)dt.
\end{eqnarray}
Indeed, if $\xi(t)$ has a white spectrum $\braket{\xi(t)\xi(t^\prime)}=\delta(t-t^\prime)$, this definition satisfies the condition~\ref{Brownian condition}
as follows.
\begin{eqnarray}
        \braket{W(t_{i+1})-W(t_i)}&=&\int^{t_{i+1}}_{t_i}\braket{\xi(t)}dt=0, \\
        \braket{(W(t_{i+1})-W(t_i))^2}&=&\int^{t_{i+1}}_{t_i}dt\int^{t_{i+1}}_{t_i}dt^\prime\braket{\xi(t)\xi(t^\prime)} \nonumber \\
        &=&\int^{t_{i+1}}_{t_i}dt\int^{t_{i+1}}_{t_i}dt^\prime\delta(t-t^\prime)=t_{i+1}-t_i.
\end{eqnarray}

Next, let us define the integral,
\begin{eqnarray}\label{ito integral}
        \int^t_0b(t^\prime)\xi(t^\prime)dt^\prime=\int^t_0b(t^\prime)dW(t^\prime),
\end{eqnarray}
to integrate a Langevin equation. Here the integrand $b(t)$ can generally depend on the past stochastic process. To define this integral,
we will approximate the integrand by a simple process at first, and then take a limit.

\begin{figure}
        \begin{center}
                \includegraphics[width=8cm]{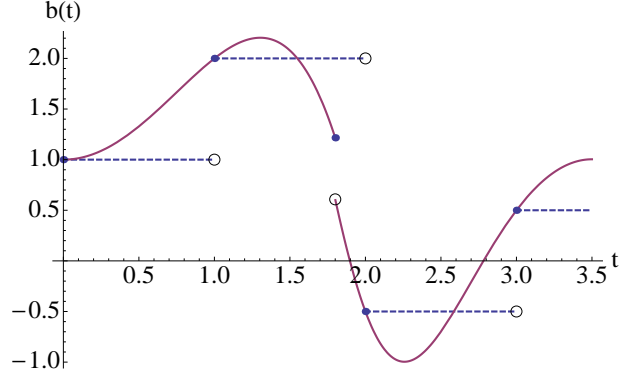}
                \caption{The approximation of an integrand. The redline represents an integrand which can depend on the past stochastic process, and
                the dashed lines are approximations of that integrand. The integrand is approximated by the initial value in each sub-period.}
                \label{integrand}
        \end{center}
\end{figure}

In the first place, $\Pi_n=\{t_0,t_1,\cdots,t_n\}$ is defined as a partition of $[0,t]$, namely
\begin{eqnarray}
        0=t_0\le t_1\le\cdots\le t_n=t.
\end{eqnarray}
Then in each sub-period $[t_i,t_{i+1})$, the integrand $b(t)$ is approximated by the constant function $b_n(t)=b(t_i)$. In other words,
$b(t)$ is approximated by the initial value in each sub-period (see also figure~\ref{integrand}). Generally, we can choose the partition $\Pi_n$ for
the approximation function $b_n(t)$ to be closer to the integrand $b(t)$ in the following sense.
\begin{eqnarray}
        \lim_{n\to\infty}\int^t_0\braket{(b_n(t)-b(t))^2}dt=0.
\end{eqnarray}  
Finally, with use of this approximation function, we define the integral~(\ref{ito integral}) as
\begin{eqnarray}
        \int^t_0b(t^\prime)dW(t^\prime)=\lim_{n\to0}\sum^{n-1}_{i=0}b_n(t_i)\left[W(t_{i+1})-W(t_i)\right].
\end{eqnarray}

The integral defined by approximating the integrand by the initial value in each sub-period like above is called ``Ito integral".
The point to notice is that the value of the stochastic integral can depend on where the integrand is approximated unlike the ordinary integral.  
The integral approximating the integrand by the midpoint value $b\left(\frac{t_i+t_{i+1}}{2}\right)$ is called ``Stratonovich integral" and its value
can be different from that of Ito integral. However, we should use Ito integral for the noise during inflation because the noise amplitude 
$\mathcal{P}_\phi(N)$ should be evaluated just before the inflaton receives the noise, otherwise the causality is broken. 

In this paper, all Langevin equations like
\begin{eqnarray}
        \dif{X}{t}=a(t)+b(t)\xi(t),
\end{eqnarray}
are defined as Ito integral,
\begin{eqnarray}
        dX(t)=a(t)dt+b(t)dW(t), \quad \text{or} \quad X(t)=X(0)+\int^t_0a(t^\prime)dt^\prime+\int^t_0b(t^\prime)dW(t^\prime).
\end{eqnarray}
In the next subsection, we will introduce the numerical integrations of Ito type.

\subsection{Numerical method}
In the numerical integration methods, the most standard one is a time discretization. For the Ito process
\begin{eqnarray}
        X(t)=X(0)+\int^t_0a(t^\prime,X(t^\prime))dt^\prime+\int^t_0b(t^\prime,X(t^\prime))dW(t^\prime),
\end{eqnarray}
the simplest finite difference approximation is given by the Euler-Maruyama method:
\begin{eqnarray}
        Y_{n+1}=Y_n+a(t_n,Y_n)\Delta_n+b(t_n,Y_n)\Delta W_n,
\end{eqnarray}
where the initial value is $Y_0=X_0$ and the step sizes are
\begin{eqnarray}
        \Delta_n=t_{n+1}-t_n, \quad \Delta W_n=W(t_{n+1})-W(t_n).
\end{eqnarray}
In the viewpoint of the numerical integration, $\Delta W_n$ is a Gaussian distributed random variable whose expectation and variance
are zero and $\Delta_n$ respectively,
\begin{eqnarray}
        \braket{\Delta W_n}=0, \quad \braket{\Delta W_n^2}=\Delta_n.
\end{eqnarray}

Next, let us mention the index of strong convergence. 
\begin{definition}
        If for the numerical approximation $Y_n,n=0,1,\cdots,N$ of a stochastic process $X(t),t\in[0,T]$, 
        there are some finite constant $K$ and positive constant
        $\delta_0$ satisfying
        \begin{eqnarray}
                \braket{|X_T-Y_N|}\le K\delta^\gamma, \quad \gamma\in(0,\infty],
        \end{eqnarray}
        for any time partition whose maximum step size is $\delta\in(0,\delta_0)$, this approximation is said to converge strongly with order $\gamma$.
\end{definition}
The convergence of stochastic process is generally bad and the order of the approximation for the stochastic differential equation is often smaller than
that for the ordinary differential equation. In fact, though the order of the Euler-Maruyama approximation for the ordinary differential equation is $1.0$,
it has been proved that the order of that for the stochastic differential equation is $0.5$.

Finally, we introduce the Runge-Kutta method for the stochastic differential equation. The Runge-Kutta method is quite practical since it can give
stable solution even with a large step size. For the Ito process depending on independent $m$-dimension Brownian motion,
\begin{eqnarray}
        X(t)=X(0)+\int^t_0a(t^\prime,X(t^\prime))dt+\sum^m_{j=1}\int^t_0b^j(t^\prime,X(t^\prime))dW^j(t^\prime),
\end{eqnarray}
The $s$-staged Runge-Kutta is parameterized as follows generally.
\begin{eqnarray}
        Y_{n+1}&=&Y_n+\sum^s_{i=1}\alpha_ia(t_n+c_i^{(0)}\Delta_n,H_i^{(0)})\Delta_n \nonumber \\
        &&+\sum^m_{k=1}\sum^s_{i=1}(\beta_i^{(1)}\Delta W^k_n+
        \beta_i^{(2)}\sqrt{\Delta_n})b^k(t_n+c_i^{(1)}\Delta_n,H_i^{(k)}),
\end{eqnarray}
where,
\begin{eqnarray}
        H_i^{(0)}&=&Y_n+\sum^s_{j=1}A_{ij}^{(0)}a(t_n+c_j^{(0)}\Delta_n,H_j^{(0)})\Delta_n+\sum^m_{l=1}\sum^s_{j=1}B_{ij}^{(0)}b^l(t_n+c_j^{(1)}\Delta_n,
        H_j^{(l)})\Delta W^l_n, \nonumber \\
        H_i^{(k)}&=&Y_n+\sum^s_{j=1}A_{ij}^{(1)}a(t_n+c_j^{(0)}\Delta_n,H_j^{(0)})\Delta_n
        +\sum^m_{l=1}\sum^s_{j=1}B_{ij}^{(1)}b^l(t_n+c_j^{(1)}\Delta_n,H_j^{(l)})
        \frac{I^{(l,k)}_n}{\sqrt{\Delta_n}}, \nonumber \\
        A^{(0)}_{ij}&=&A^{(1)}_{ij}=B^{(0)}_{ij}=B^{(1)}_{ij}=0, \quad \text{for $i\le j$,}
\end{eqnarray}
and $A^{(0)},A^{(1)},B^{(0)},B^{(1)},c^{(0)},c^{(1)},\alpha,\beta^{(1)}$ and $\beta^{(2)}$ are the method parameters. 
$I^{(l,k)}_n$ denotes the multiple Ito integral, and when noise correlations are represented as
\begin{eqnarray}
        \braket{\xi^l(t)\xi^k(t^\prime)}=C^{lk}\delta(t-t^\prime),
\end{eqnarray}
it is written as
\begin{eqnarray}
        I^{(l,k)}_n=\frac{1}{2}\left(\Delta W^l_n\Delta W^k_n-C^{lk}\Delta_n\right).
\end{eqnarray}
Especially for independent noise, $C^{lk}$ reads the Kronecker delta $\delta^{lk}$.
The parameters are often listed with use of the extended Butcher table~\ref{Butcher}. In table~\ref{3 step 1st order},
we show the parameters of the 3-staged order 1.0 strong Runge-Kutta~\cite{Rossler} which we use in this paper.

\begin{table}
        \begin{center}
                \caption{The extended Butcher table.}
                \label{Butcher}
                \begin{tabular}{c|c|c|c}
                        \\
                        $c^{(0)}$ & $A^{(0)}$ & $B^{(0)}$ & \\ \cline{1-3}
                        $c^{(1)}$ & $A^{(1)}$ & $B^{(1)}$ & \\ \hline
                        & $\alpha^T$ & $\beta^{(1)T}$ & $\beta^{(2)T}$
                \end{tabular}
        \end{center}
\end{table}     

\begin{table}
        \begin{center}
                \caption{The parameters of the 3-staged order 1.0 strong Runge-Kutta. In this paper, we use this method.}
                \label{3 step 1st order}
                \begin{tabular}{c|ccc|ccc|ccc}
                        $0$ &&&&&&&&& \\
                        $0$ & $0$ &&& $0$ &&&&& \\
                        $0$ & $0$ & $0$ && $0$ & $0$ &&&& \\ \cline{1-7}
                        $0$ &&&&&&&&& \\
                        $0$ & $0$ &&& $1$ &&&&& \\
                        $0$ & $0$ & $0$ && $-1$ & $0$ &&&& \\ \hline
                        & $1$ & $0$ & $0$ & $1$ & $0$ & $0$ & $0$ & $1/2$ & $-1/2$
                \end{tabular}
        \end{center}
\end{table}

\section{Mixing}\label{mixing}
In this paper, the solutions with the constant masses~(\ref{const mass approx}), assuming the effect of the mass change during sub-horizon
is negligible. However, the mixing term $V_{IJ}$ remains even in this case. Though we neglect this term in this paper, the effect of mixing
can be taken as follows.

Here we treat $V_{IJ}$ as a constant and evaluate it at horizon crossing $k=\epsilon aH$
for each mode, even though it varies as time goes on actually. Taking the diagonalizing matrix $P$ of $V_{IJ}$:
\begin{eqnarray}
        \left(
        \begin{array}{ccc}
                \lambda_1 & & \\
                & \lambda_2 & \\
                & & \ddots
        \end{array}
        \right)=P^{-1}V_{IJ}P,
\end{eqnarray}
the field $P^{-1}_{IJ}\phi_{\mathbf{k}}^J$ has no mixing. Therefore, the correlator of the original field reads,
\begin{eqnarray}
        \braket{\phi_\mathbf{k}^{I\dagger}\phi_\mathbf{k}^J}
        =\braket{(P_{IL}P^{-1}_{LM}\phi_\mathbf{k}^M)^\dagger(P_{JN}P^{-1}_{NP}\phi_\mathbf{k}^P)}
        =P_{JL}\braket{(P^{-1}\phi_\mathbf{k})^{L\dagger}(P^{-1}\phi_\mathbf{k})^L}P^\dagger_{LI},
\end{eqnarray}
and the following power spectrum should be used as the amplitude of noise $\xi_I$.
\begin{eqnarray}\label{mixed power}
        \mathcal{P}_{\phi^I}=\sum_LP_{IL}\left(\frac{H^2}{8\pi}\epsilon^3|H_{\nu_L}^{(1)}(\epsilon)|^2\right)P^\dagger_{LI}.
\end{eqnarray}
Note that the summation with respect to $I$ is not taken. Noise $\xi_I$ also has non-zero correlation,
\begin{eqnarray}
        \braket{\xi_I(N)\xi_J(N^\prime)}=\frac{\sum_LP_{IL}\left(\frac{H^2}{8\pi}\epsilon^3|H_{\nu_L}^{(1)}(\epsilon)|^2\right)P^\dagger_{LI}}
        {\mathcal{P}_{\phi^I}^{1/2}\mathcal{P}_{\phi^J}^{1/2}}\delta(N-N^\prime),
\end{eqnarray}
for $I\ne J$. Here $\mathcal{P}_{\phi^I}$ denotes the power spectrum given by eq.~(\ref{mixed power}).

\end{document}